\documentclass[twocolumn]{aastex63}

\usepackage{amsmath,amstext}
\usepackage{booktabs}

\shorttitle{AT\,2019qyl: Nova in NGC\,300}
\shortauthors{Jencson et al.}

\begin{document}

\title{AT\,2019qyl in NGC 300: Internal Collisions in the Early Outflow from a Very Fast Nova in a Symbiotic Binary\footnote{This paper includes data gathered with the 6.5~m Magellan Telescopes located at Las Campanas Observatory, Chile.}\footnote{Based on observations made with the NASA/ESA \textit{Hubble Space Telescope}, obtained from the data archive at the Space Telescope Science Institute. STScI is operated by the Association of Universities for Research in Astronomy, Inc., under NASA contract NAS 5-26555.}}

\correspondingauthor{Jacob E. Jencson}
\email{jjencson@email.arizona.edu}

\author[0000-0001-5754-4007]{Jacob E. Jencson}
\affil{Steward Observatory, University of Arizona, 933 North Cherry Avenue, Tucson, AZ 85721-0065, USA}

\author[0000-0003-0123-0062]{Jennifer E. Andrews}
\affil{Steward Observatory, University of Arizona, 933 North Cherry Avenue, Tucson, AZ 85721-0065, USA}

\author[0000-0003-1377-7145]{Howard E. Bond}  
\affil{Department of Astronomy \& Astrophysics, Pennsylvania State University, University Park, PA 16802, USA}
\affil{Space Telescope Science Institute, 3700 San Martin Drive, Baltimore, MD 21218, USA}

\author{Viraj Karambelkar}
\affiliation{Division of Physics, Mathematics and Astronomy, California Institute of Technology, Pasadena, CA 91125, USA}

\author[0000-0003-4102-380X]{David J. Sand}
\affil{Steward Observatory, University of Arizona, 933 North Cherry Avenue, Tucson, AZ 85721-0065, USA}

\author[0000-0001-9038-9950]{Schuyler D.\ van Dyk}
\affil{IPAC, California Institute of Technology, 1200 E. California Boulevard, Pasadena, CA 91125, USA}

\author[0000-0003-0901-1606]{Nadejda Blagorodnova}
\affil{Department of Astrophysics/IMAPP, Radboud University, Nijmegen, The Netherlands}

\author[0000-0003-4850-9589]{Martha L. Boyer}
\affil{Space Telescope Science Institute, 3700 San Martin Drive, Baltimore, MD 21218, USA}

\author{Mansi M. Kasliwal}
\affiliation{Division of Physics, Mathematics and Astronomy, California Institute of Technology, Pasadena, CA 91125, USA}

\author{Ryan M. Lau}
\affil{Institute of Space \& Astronautical Science, Japan Aerospace Exploration Agency, 3-1-1 Yoshinodai, Chuo-ku, Sagamihara, Kanagawa
252-5210, Japan}

\author[0000-0002-1856-9225]{Shazrene Mohamed}
\affil{South African Astronomical Observatory, P.O.\ Box 9, 7935 Observatory, South Africa}
\affil{Department of Astronomy, University of Cape Town, Private Bag X3, Rondebosch 7701, South Africa}
\affil{National Institute for Theoretical Physics (NITheP), KwaZulu-Natal, South Africa}

\author{Robert Williams}
\affiliation{Department of Astronomy and Astrophysics, University of California, Santa Cruz, CA 95064, USA}
\affiliation{Space Telescope Science Institute, 3700 San Martin Drive, Baltimore, MD 21218, USA}

\author[0000-0002-4678-4432]{Patricia A. Whitelock}
\affil{South African Astronomical Observatory, P.O.\ Box 9, 7935 Observatory, South Africa}
\affil{Department of Astronomy, University of Cape Town, Private Bag X3, Rondebosch 7701, South Africa}

\author[0000-0002-1546-9763]{Rachael C. Amaro}
\affil{Steward Observatory, University of Arizona, 933 North Cherry Avenue, Tucson, AZ 85721-0065, USA}

\author[0000-0002-4924-444X]{K. Azalee Bostroem}
\affil{Department of Physics and Astronomy, University of California, 1 Shields Avenue, Davis, CA 95616-5270, USA}

\author[0000-0002-7937-6371]{Yize Dong}
\affil{Department of Physics and Astronomy, University of California, 1 Shields Avenue, Davis, CA 95616-5270, USA}

\author{Michael J. Lundquist}
\affil{Steward Observatory, University of Arizona, 933 North Cherry Avenue, Tucson, AZ 85721-0065, USA}

\author[0000-0001-8818-0795]{Stefano Valenti}
\affil{Department of Physics and Astronomy, University of California, 1 Shields Avenue, Davis, CA 95616-5270, USA}

\author[0000-0003-2732-4956]{Samuel D. Wyatt}
\affil{Steward Observatory, University of Arizona, 933 North Cherry Avenue, Tucson, AZ 85721-0065, USA}

\author[0000-0003-0035-6659]{Jamie Burke}
\affil{Department of Physics, University of California, Santa Barbara, CA 93106-9530, USA}
\affil{Las Cumbres Observatory, 6740 Cortona Drive, Suite 102, Goleta, CA 93117-5575, USA}

\author{Kishalay De}
\affiliation{Division of Physics, Mathematics and Astronomy, California Institute of Technology, Pasadena, CA 91125, USA}

\author[0000-0001-8738-6011]{Saurabh W. Jha}
\affil{Department of Physics and Astronomy, Rutgers, the State University of New Jersey, 136 Frelinghuysen Road, Piscataway, NJ 08854, USA}

\author[0000-0001-5975-290X]{Joel Johansson}
\affil{The Oskar Klein Centre, Department of Physics, AlbaNova, Stockholm University, SE-10691 Stockholm, Sweden}

\author[0000-0002-7559-315X]{C\'{e}sar Rojas-Bravo}
\affil{Department of Astronomy and Astrophysics, University of California, Santa Cruz, CA 95064, USA}

\author{David A. Coulter}
\affil{Department of Astronomy and Astrophysics, University of California, Santa Cruz, CA 95064, USA}

\author{Ryan J. Foley}
\affil{Department of Astronomy and Astrophysics, University of California, Santa Cruz, CA 95064, USA}

\author{Robert D. Gehrz}
\affil{Minnesota Institute for Astrophysics, School of Physics and Astronomy, University of Minnesota, 116 Church Street SE, Minneapolis, MN 55455, USA}

\author{Joshua Haislip}
\affil{Department of Physics and Astronomy, University of North Carolina at Chapel Hill, Chapel Hill, NC 27599, USA}

\author[0000-0002-1125-9187]{Daichi Hiramatsu}
\affil{Department of Physics, University of California, Santa Barbara, CA 93106-9530, USA}
\affil{Las Cumbres Observatory, 6740 Cortona Drive, Suite 102, Goleta, CA 93117-5575, USA}

\author{D. Andrew Howell}
\affil{Department of Physics, University of California, Santa Barbara, CA 93106-9530, USA}
\affil{Las Cumbres Observatory, 6740 Cortona Drive, Suite 102, Goleta, CA 93117-5575, USA}

\author{Charles D. Kilpatrick}
\affil{Department of Astronomy and Astrophysics, University of California, Santa Cruz, CA 95064, USA}

\author[0000-0002-8532-9395]{Frank J. Masci}
\affil{IPAC, California Institute of Technology, 1200 E. California Boulevard, Pasadena, CA 91125, USA}

\author{Curtis McCully}
\affil{Department of Physics, University of California, Santa Barbara, CA 93106-9530, USA}
\affil{Las Cumbres Observatory, 6740 Cortona Drive, Suite 102, Goleta, CA 93117-5575, USA}

\author[0000-0001-8771-7554]{Chow-Choong Ngeow}
\affil{Graduate Institute of Astronomy, National Central University, 300 Jhongda Road, 32001 Jhongli, Taiwan}

\author{Yen-Chen Pan}
\affil{Graduate Institute of Astronomy, National Central University, 300 Jhongda Road, 32001 Jhongli, Taiwan}

\author{Craig Pellegrino}
\affil{Department of Physics, University of California, Santa Barbara, CA 93106-9530, USA}
\affil{Las Cumbres Observatory, 6740 Cortona Drive, Suite 102, Goleta, CA 93117-5575, USA}

\author{Anthony L. Piro}
\affil{The Observatories of the Carnegie Institution for Science, 813 Santa Barbara Street, Pasadena, CA 91101, USA}

\author[0000-0003-3642-5484]{Vladimir Kouprianov}
\affil{Department of Physics and Astronomy, University of North Carolina at Chapel Hill, Chapel Hill, NC 27599, USA}
\affil{Central (Pulkovo) Observatory of the Russian Academy of Sciences, 196140, 65/1 Pulkovskoye Avenue, Saint Petersburg, Russia}

\author[0000-0002-5060-3673]{Daniel E. Reichart}
\affil{Department of Physics and Astronomy, University of North Carolina at Chapel Hill, Chapel Hill, NC 27599, USA}

\author[0000-0002-4410-5387]{Armin Rest}
\affil{Space Telescope Science Institute, 3700 San Martin Drive, Baltimore, MD 21218, USA}
\affil{The Johns Hopkins University, Baltimore, MD 21218, USA}

\author{Sofia Rest}
\affil{The Johns Hopkins University, Baltimore, MD 21218, USA}

\author{Nathan Smith}
\affil{Steward Observatory, University of Arizona, 933 North Cherry Avenue, Tucson, AZ 85721-0065, USA}

\begin{abstract}
Nova eruptions, thermonuclear explosions on the surfaces of white dwarfs (WDs), are now recognized to be among the most common shock-powered astrophysical transients. We present the early discovery and rapid ultraviolet (UV), optical, and infrared (IR) temporal development of AT\,2019qyl, a recent nova in the nearby Sculptor Group galaxy NGC\,300. The light curve shows a rapid rise lasting $\lesssim 1$\,day, reaching a peak absolute magnitude of $M_V = -9.2$\,mag, and a very fast decline, fading by 2~mag over 3.5\,days. A steep dropoff in the light curves after 71\,days and the rapid decline timescale suggest a low-mass ejection from a massive WD with $M_{\rm WD} \gtrsim 1.2~M_{\odot}$. We present an unprecedented view of the early spectroscopic evolution of such an event. Three spectra prior to the peak reveal a complex, multicomponent outflow giving rise to internal collisions and shocks in the ejecta of an He/N-class nova. We identify a coincident IR-variable counterpart in the extensive preeruption coverage of the transient location and infer the presence of a symbiotic progenitor system with an O-rich asymptotic-giant-branch donor star, as well as evidence for an earlier UV-bright outburst in 2014. We suggest that AT\,2019qyl is analogous to the subset of Galactic recurrent novae with red-giant companions such as RS Oph and other embedded nova systems like V407 Cyg. Our observations provide new evidence that \replaced{internal outflow collisions}{internal shocks between multiple, distinct outflow components}
likely \replaced{play an important role in}{contribute to the generation of} the shock-powered emission from such systems.
\end{abstract}

\keywords{Novae (1127); Symbiotic binary stars (1674); Recurrent novae (1366); White dwarf stars (1799); Asymptotic giant branch stars (2100); Spectroscopy (1558);}

\section{Introduction} \label{sec:intro}
Novae are a class of cataclysmic variables (CVs) whose eruptions are the result of a thermonuclear runaway (TNR) on the surface of a white dwarf (WD) accreting hydrogen-rich material from a nondegenerate companion \citep{gallagher78}. Novae are among the most common explosive thermonuclear transients \citep[e.g.,][]{darnley06,shafter17,de21} and greatly contribute to the chemical enrichment of galaxies (e.g., \citealp{gehrz88a}, and see \citealp{gehrz98,jose06} for reviews). Nova eruptions are also unique probes of the underlying CV population, which is key to understanding mass transfer in binaries across a range of masses and evolutionary stages \citep[e.g.,][]{townsley05,nelemans16}. 

While novae have been studied in earnest for more than a century \citep[e.g.,][]{ritchey17,payne-gaposchkin57}, our understanding of their eruptions remains incomplete. The last decade has witnessed a substantial renewed interest in complex mass ejections and internal shocks in multicomponent nova outflows (see \citealp{chomiuk20} for a recent review) beginning with the discovery of gamma-ray emission from novae by NASA's \textit{Fermi Gamma-ray Space Telescope} \citep{abdo10,ackermann14,cheung16}. Building on prior lines of evidence, recent observations, including correlations between gamma-ray and optical light-curve peaks \citep{metzger14,li17,aydi20b}; hard X-rays ($\gtrsim$1~keV) arising from hot, shock-heated plasma \citep[e.g.,][]{lloyd92,orio01,mukai08,gordon21}; and nonthermal radio synchrotron emission \citep[e.g.,][]{taylor87,chomiuk14,finzell18}, have increasingly demonstrated the importance of shocks in powering nova emission. 

Spectroscopic evidence for multiphase outflows has also been seen in the evolution of optical emission lines, particularly during the earliest phases of nova eruptions \citep{mclaughlin42,gallagher78}. \citet{aydi20a} revisited this topic with a sample of premaximum spectra of 12 novae, all showing consistent evidence for distinct velocity components in their ejecta. This scenario is expected to invariably give rise to shocks \citep{mclaughlin47,friedjung87,friedjung11,friedjung93} and provides important clues on the yet poorly understood mass-loss mechanisms that may operate in nova eruptions. Their sample was limited to classical novae (CNe; typically occurring in close binary systems harboring a WD and a main-sequence companion that overflows its Roche lobe) and notably did not include examples of the fastest-evolving novae or systems in wide binaries with red-giant (RG) donor stars. 

Some RG novae are of particular interest as the progenitors of a subclass of recurrent novae (RNe; nova systems historically observed in outburst at least twice) similar to the famed RS Oph, whose previous 2006 eruption was exceptionally well studied \citep[see, e.g.,][and references therein]{evans08}. The known sample of RNe in the Milky Way currently stands at 10 objects, with the latest addition being V2487 Oph (\citealp{pagnotta09}; see \citealp{schaefer10a} for a comprehensive review). They all have recurrence times $< 100$~yr \citep{schaefer10a}, although this is almost certainly a selection effect based on the time for which reliable astronomical records are available. Essentially all novae are expected to recur, with the predicted recurrence times for some systems exceeding $10^6$~yr \citep{ford78,yaron05,wolf13}. 

The recurrence time of a nova is set by the time to build up sufficient accreted mass to trigger the TNR---primarily a function of the accretion rate and the WD mass ($M_{\mathrm{WD}}$). The shortest recurrence times tend to occur on massive WDs with lower critical envelope masses owing to their larger surface gravities and pressures. Indeed, \replaced{the}{some} RNe for which WD masses have been reliably measured are near the Chandrasekhar mass ($\gtrsim$1.2~$M_{\odot}$; \added{including RS~Oph and V745~Sco}; \citealp{osborne11,page15}). \added{A notable exception is T~Pyx, whose recurrent eruptions are believed to arise on a $\approx$1~$M_{\odot}$ WD with a somewhat inflated main-sequence companion \citep{uthas10,nelson14}.} \replaced{As such, RNe}{RNe on massive WDs} have long been considered possible progenitors of Type~Ia supernovae \citep[SNe; e.g.,][]{starrfield85,dellavalle96,schaefer10a,kato12,maoz14}, especially the extragalactic population of rapid recurrent novae discovered in recent years \citep{darnley20}. Systems like the remarkable M31N~2008-12a \citep{darnley14,tang14,henze15a,henze15b,darnley17}, with a recurrence timescale $\lesssim 1$\,yr, may be due for a  thermonuclear SN explosion within $10^6$\,yr \citep{kato14}.

Owing to their low envelope masses at the time of explosion, RNe are also among the fastest-evolving novae. Thus, obtaining early observations of RNe that can reveal the kinematics of the outflow and mass-loss mechanisms is particularly challenging. Additional difficulties arise for RG novae, where the outburst is embedded in the dense wind of the companion, leading to external interactions that quickly dominate the spectral evolution. Opportunities to observe Galactic RNe are limited by the small, and slowly growing, sample of known systems. Outside the Milky Way, a total of 4 recurrent novae have now been identified in the LMC, and there are 18 known systems in M31 \citep[][and references therein]{darnley20}. Modern, high-cadence transient surveys are well equipped to discover fast and faint transients in galaxies beyond the Local Group and offer the most expedient route to obtaining the early (within hours of eruption) observations necessary to probe the outflow structure and mass-loss mechanisms of embedded RG novae and rapidly evolving, extragalactic analogs to RNe in the Milky Way.

Here we present the discovery of the recent nova AT\,2019qyl in the nearby Sculptor Group galaxy NGC\,300, and identify its luminous RG counterpart. In Section~\ref{sec:observations}, we describe the discovery and intensive follow-up observations, including multiple epochs of very early imaging and spectroscopy during the first day of the outburst. In Section~\ref{sec:transient}, we describe the host environment and evolution of the transient, including our multiband ultraviolet (UV), optical, and infrared (IR) light curves covering the first 127 days (Section~\ref{sec:lcs}) and the optical and near-IR (NIR) spectroscopic sequence (Section~\ref{sec:spectral_evolution}). Section~\ref{sec:line_profiles} presents a detailed look at the profiles of several strong emission lines and examines the kinematics of the ejecta.  Section~\ref{sec:arc_im} summarizes the properties of the progenitor system and its variability inferred from extensive multiwavelength (UV to IR) space and ground-based archival imaging. Section~\ref{sec:RNe} places AT\,2019qyl in the context of Galactic RNe and other embedded RG novae including V407 Cyg. In Section~\ref{sec:shocks}, we discuss the evidence for, and origins of, multiple, distinct episodes of mass ejection during the nova outburst and the resulting collisions and shocks. Section~\ref{sec:summary} offers a brief summary of our results and main conclusions. 

\section{Discovery and Follow-up Observations} \label{sec:observations}
\subsection{DLT40 Discovery in NGC\,300}\label{sec:disc}
AT\,2019qyl was discovered on UT 2019 September 26.21 (MJD 58752.21) by the Distance $<$ 40\,Mpc subday-cadence supernova (SN) search (DLT40; see \citealp{tartaglia18} for survey details and \citealp{bostroem20} for recent improvements in our transient detection and triggering algorithms).  The unfiltered discovery data were taken with the 0.4\,m Panchromatic Robotic Optical Monitoring and Polarimetry Telescopes (PROMPT) PROMPT5 telescope at the Cerro Tololo Inter-American Observatory (CTIO) operated by the Skynet telescope network \citep{reichart05}. Its discovery magnitude was $17.4$\,mag, calibrated to the AAVSO Photometric All-Sky Survey (APASS;\footnote{\url{https://www.aavso.org/apass}} \citealp{henden14}) $r$ band. An earlier DLT40 nondetection at $\approx$19.0\,mag on 2019 September 25.22 constrains the time of eruption to within 0.99\,days of the first detection. The source was internally named DLT\,19m by the DLT40 team and reported to the Transient Name Server where it received the official IAU designation AT\,2019qyl \citep{valenti19}. A subsequent, independent detection (ATLAS-c ``cyan'' filter) was reported on 2019 September 27.50 by the ATLAS survey \citep{tonry18}, along with an earlier nondetection (ATLAS-o ``orange'' filter) to a limiting magnitude of $19.4$\,mag on 2019 September 25.46, further constraining the time of eruption to within 0.75\,days of the first DLT40 detection. 

After the initial discovery and confirmation of AT~2019qyl, we took a sequence of PROMPT5 images of the field over the next 4.3~hr from CTIO, during which the transient brightened from the initial $r=17.4$ mag to 17.2 mag (see\deleted{ inset of Figure~\ref{fig:lcs}, and} further discussion in Section~\ref{sec:t0}). The DLT40 team also triggered a rapid-response program with the Neil Gehrels {\it Swift} Observatory \citep[hereafter {\it Swift};][]{gehrels04} in place to acquire high-cadence early UV light curves of nearby transients, with the first data arriving $\approx$4\,hr after discovery. Immediate follow-up spectroscopy through a Gemini rapid target-of-opportunity (ToO) program ($\approx$2\,hr after discovery, see Section~\ref{sec:spec_obs}) led to an initial classification report \citep{andrews19} that misclassified the transient as an outburst of a luminous blue variable star or possible young Type IIn SN based on the presence of narrow/intermediate-width emission features of H and \ion{He}{1}. However, as described in more detail in Section~\ref{sec:spectral_evolution}, our spectroscopic analysis confirms this source as a nova. 

\begin{figure*}
\includegraphics[width=\textwidth]{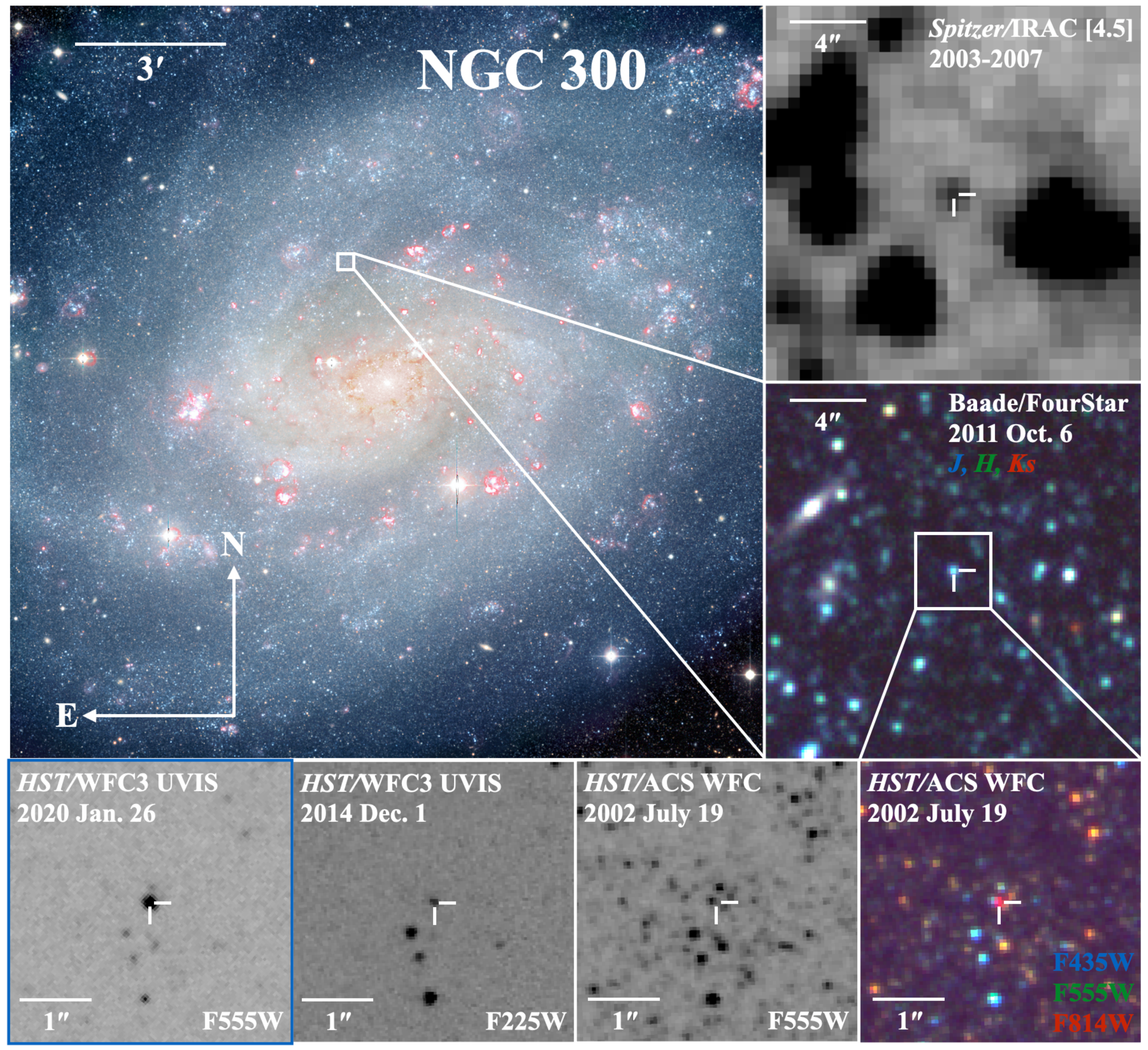}
\caption{\label{fig:pre-imaging}
Pre- and postoutburst imaging of the site of AT\,2019qyl in NGC\,300. The upper, leftmost panel shows an archival, color-composite image of the host ($B$ in blue, $V$ in green, and $R$ and H$\alpha$ in red; credit: MPG/ESO). The location of AT\,2019qyl along a northern spiral arm of the galaxy is shown in more detail in the $20\arcsec \times 20\arcsec$ middle, rightmost zoom-in panel based on Baade/FourStar NIR images of NGC\,300 from 2011 October 6 ($J$ in blue, $H$ in green, and $K_s$ in red). There is a relatively isolated point source consistent with the position of AT\,2019qyl detected in all three NIR filters. Above in the top right corner, we show the same region in the archival \textit{Spitzer}/IRAC [4.5] Super Mosaic (2003--2007 stack), where we also identify a coincident mid-IR (MIR) precursor source. The bottom rightmost panel shows a zoom-in view of the immediate $4\arcsec \times 4\arcsec$ region around the transient in the archival \textit{HST} ACS/WFC images from 2002 July 19 in three filters (F435W in blue, F555W in green, and F814W in red). At the precise location of the transient, determined from our 2020 January 26 follow-up \textit{HST}/WFC3 UVIS observations in F555W (bottom, leftmost panel in blue outline), we identified a very red source in the archival \textit{HST} frames detected in F814W, but not in F555W (bottom, center-right panel) or F435W. In the WFC3/UVIS F225W image from 2014 December 1 (bottom, center-left panel) we also identify a UV point source consistent with the transient position, approximately five years before the nova eruption. 
}
\end{figure*}

At a position of $00^{\mathrm{h}}54^{\mathrm{m}}57\fs680, -37\degr38\arcmin40\farcs01$ (J2000.0), the transient is located in a northern spiral arm of the nearby star-forming galaxy NGC\,300 and 2\farcm54 from the galaxy's center (see location in Figure~\ref{fig:pre-imaging}). 
Throughout this work, we assume a distance modulus for NGC\,300 of $(m-M)_0 = 26.29 \pm 0.07$ ($D=1.81$\,Mpc), based on the most recently available Cepheid measurement by \citet{bhardwaj16}. We adopt a value for the Galactic extinction toward NGC\,300 of $E(B-V) = 0.01$\,mag, based on the \citet{schlafly11} recalibration of the \citet{schlegel98} dust maps and assume a standard \citep{fitzpatrick99} reddening law with $R_V = 3.1$.

\subsection{Imaging Observations}\label{sec:im_obs}
We obtained a sequence of images with the DLT40 PROMPT5 0.4-m telescope at CTIO. The PROMPT5 telescope has no filter (`Open'), which we calibrate to the $r$ band \citep[see][for further reduction details]{tartaglia18}. Immediately after discovery, we began an intense photometric campaign with the Las Cumbres Observatory global telescope network (\citealp{brown13}) in the $UBVgri$ bands with the Sinistro cameras on the 1 m telescopes at CTIO (Chile), Siding Spring (Australia), and Sutherland (South Africa). These data were taken as part of the Global Supernova Project (GSP), as well as PI-led programs ($griz_s$; OPTICON 19B-053; PI N.~Blagorodnova). The images were reduced with  the Beautiful Algorithms to Normalize Zillions of Astronomical Images (BANZAI) pipeline \citep{mccully18}. For GSP data, we performed point-spread function (PSF)-fitting photometry without template subtraction using {\sc lcogtsnpipe} \citep{valenti16}, a PyRAF-based reduction pipeline. $BV$- and $gri$-band data were calibrated to Vega and AB magnitudes, respectively, using APASS DR9 \citep{henden16} catalog stars in the images. For $U$-band data, we used Landolt standard-star observations taken on the same night by the same telescope. For the additional Las Cumbres data in the $griz_s$ bands, aperture photometry was performed using a custom Python-based pipeline$\footnote{\url{https://github.com/nblago/utils/tree/master/src/photometry}}$ and calibrated to the SkyMapper catalog \citep{wolf_skymapper13}. As the source is relatively isolated in NGC\,300, both methods of photometry give very similar results.

The Swope 1 m telescope at Las Campanas Observatory (LCO) was used for $uBVgri$ observations with the Direct 4k$\times$4k imager. All bias subtraction, flat fielding, image stitching, registration, and photometric calibration were performed using {\sc photpipe} \citep{rest05} as described in \citet{kilpatrick18b}.  Similarly, $BVgri$ imaging was taken with the Lulin One-meter telescope in Taiwan, and was reduced with standard IRAF tasks; final image calibration was done utilizing the Swope reduction pipeline. For these images, aperture photometry was performed for AT\,2019qyl on the reduced images without image differencing. 

UV and optical images were obtained during the early portion of the light curve with the Ultraviolet/Optical telescope (UVOT; \citealp{roming05}) on board \textit{Swift}. The data were downloaded from the NASA \textit{Swift} Data Archive\footnote{\url{https://heasarc.gsfc.nasa.gov/cgi-bin/W3Browse/swift.pl}}, and the images were reduced using standard software distributed with \texttt{HEAsoft}\footnote{\url{https://heasarc.gsfc.nasa.gov/docs/software/heasoft/}}. Photometry was performed for all the $uvw1$, $uvm2$, $uvw2$, $U$-, $B$-, and $V$-band images using a 3\farcs0 aperture at the location of AT\,2019qyl. We subtracted the contribution from the host galaxy using an identical aperture on available preoutburst imaging in all bands. 

The location of AT\,2019qyl was multiply imaged with the Infrared Array Camera (IRAC; \citealp{fazio04}) on board the \textit{Spitzer Space Telescope} \citep{werner04,gehrz07} in the 3.6 and 4.5~$\mu$m imaging channels ([3.6] and [4.5]) between 2014 and the end of 2019 during regular monitoring of NGC\,300 by the SPitzer Infrared Intensive Transients Survey (SPIRITS; PI: M.\ Kasliwal; PIDs 10136, 11063, 13053, 14089) and in observations targeting the ultraluminous X-ray source NGC\,300~ULX1 (PI: R.\ Lau; PID 14270). The postbasic calibrated data level images were downloaded from the \textit{Spitzer} Heritage Archive\footnote{\url{https://sha.ipac.caltech.edu/applications/Spitzer/SHA/}} and \textit{Spitzer} Early Release Data Service\footnote{\url{http://ssc.spitzer.caltech.edu/warmmission/sus/mlist/archive/2015/msg007.txt}} and processed through an automated image-subtraction pipeline (for survey and pipeline details, see \citealp{kasliwal17,jencson19}). For reference images, we used the Super Mosaics,\footnote{Super Mosaics are available as \textit{Spitzer} Enhanced Imaging Products through the NASA/IPAC Infrared Science Archive: \url{https://irsa.ipac.caltech.edu/data/SPITZER/Enhanced/SEIP/overview.html}} consisting of stacks of images obtained between 2003 November 21 and 2007 December 29. We performed aperture photometry on our difference images adopting the appropriate aperture corrections from the IRAC instrument handbook\footnote{\url{http://irsa.ipac.caltech.edu/data/SPITZER/docs/irac/iracinstrumenthandbook/}} and following the method for a robust estimate of the photometric uncertainties as described in \citet{jencson20}. We converted our flux measurements to Vega-system magnitudes using the zero-magnitude fluxes presented for
each IRAC channel in the IRAC instrument handbook.

We executed ToO observations with the 
\textit{Hubble Space Telescope} (\textit{HST}) Wide Field Camera 3 (WFC3) UVIS channel in subarray mode in F555W (23 frames, 690\,s total
exposure; PI S.\ Van Dyk; PID GO-15151) on 2020 January 26.06 with the primary goal of obtaining a precise position for AT\,2019qyl in comparison with archival imaging (see Section~\ref{sec:arc_im} for details). To obtain photometry of AT\,2019qyl, we processed the data with \texttt{Dolphot} \citep{dolphin00,dolphin16}, first running the individual frames corrected for charge transfer efficiency through \texttt{AstroDrizzle} \citep{astrodrizzle12}, to flag cosmic-ray hits. We adopt a weighted average of the \texttt{Dolphot} measurements from 21 frames of $\mathrm{F555W} = 22.52 \pm 0.07$\,mag (Vega). 

\begin{figure*}
\centering
\includegraphics[width=0.7\textwidth]{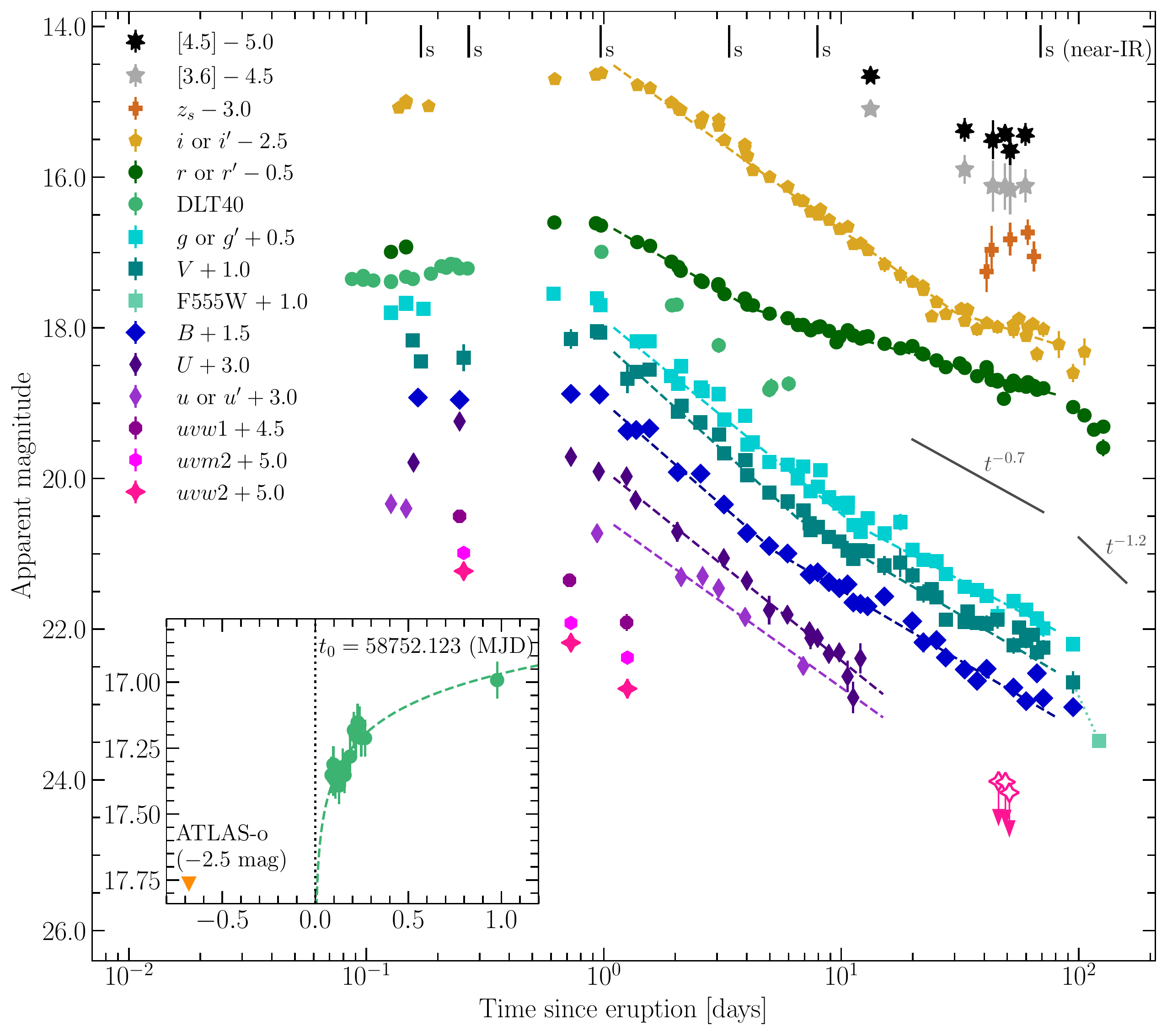}
\caption{\label{fig:lcs}
The postexplosion, multiband light curves of AT\,2019qyl in the UV, optical, and IR. The times corresponding to our spectroscopic observations reported in Table~\ref{table:spec} are indicated by the black ticks labeled with an ``s''. Broken power-law fits to flux measurements are shown as the dashed curves in the corresponding colors for each of the $UBV$ and $ugri$ light curves. The best-fitting parameters defining these curves are reported in Table~\ref{table:lc_properties}. A dotted curve between the last $V$-band measurement and the late-time F555W measurement is shown to illustrate the steep dropoff in the light curve occurring around $t\approx80$~days. Power-law declines corresponding to the free-free emission model light curves of \citet{hachisu06a,hachisu07} are shown for comparison as black solid curves. In the bottom-left inset, we show the early-time DLT40 light curve of AT\,2019qyl during the rise of the transient, with time $t$ since $t_0$ in days on the $x$-axis (linear scale), and the most constraining preeruption nondetection from ATLAS (ATLAS-o$ - 2.5$\,mag, orange triangle). Based on a power-law fit to the early data (green dashed curve), we adopt an explosion time of 2019 September 26.12 (MJD 58752.12). All broadband photometry is corrected for Milky Way extinction assuming $E(B-V) = 0.01$. 
}
\end{figure*}

We show our multiband light curves of the transient in Figure~\ref{fig:lcs}, including only measurements with errors of $<$0.2\,mag. Our complete set of photometric measurements is provided in machine-readable format in Table~\ref{table:phot}. 

\begin{deluxetable*}{ccccc}
\tablecaption{Photometry from Follow-up Observations \label{table:phot}}
\tablehead{\colhead{MJD} & \colhead{Phase\tablenotemark{a}} & \colhead{Tel/Inst.} & \colhead{Band} & \colhead{App.\ Magnitude\tablenotemark{b}} \\ 
\colhead{} & \colhead{(days)} & \colhead{} & \colhead{} & \colhead{(mag)} }
\startdata
58752.21 &   0.09 & DLT40                     & Open & $17.38$ $(0.07)$ \\
58752.22 &   0.10 & DLT40                     & Open & $17.34$ $(0.07)$ \\
58752.22 &   0.10 & DLT40                     & Open & $17.39$ $(0.07)$ \\
58752.23 &   0.11 & DLT40                     & Open & $17.40$ $(0.07)$ \\
58752.25 &   0.13 & DLT40                     & Open & $17.41$ $(0.07)$ \\
58752.25 &   0.13 & DLT40                     & Open & $17.42$ $(0.07)$ \\
58752.25 &   0.13 & Las Cumbres 1\,m          & $r'$             & $17.52$ $(0.01)$ \\
58752.25 &   0.13 & Las Cumbres 1\,m          & $u'$             & $17.39$ $(0.03)$ \\
58752.25 &   0.13 & Las Cumbres 1\,m          & $g'$             & $17.34$ $(0.01)$ \\
58752.26 &   0.14 & Las Cumbres 1\,m          & $i'$             & $17.60$ $(0.01)$ \\
\enddata
\tablenotetext{a}{Phase refers to time since $t_0$ on MJD 58752.12.}
\tablenotetext{b}{1$\sigma$ uncertainties are given in parentheses. Ground-based magnitudes are given in their native system, Vega magnitudes for $UBV$ and AB magnitudes for $ugri$ and $u'g'r'i'z_s$. DLT40 instrumental magnitudes are calibrated to $r$-band in AB magnitudes. For space-based facilities (\textit{HST}, \textit{Spitzer}, and \textit{Swift}), measurements are in the Vega system.}
\tablecomments{Table 1 is published in its entirety in the machine-readable format. A portion is shown here for guidance regarding its form and content.}
\end{deluxetable*}


\subsection{Spectroscopy}\label{sec:spec_obs}
\begin{figure*}
\centering
\includegraphics[width=\textwidth]{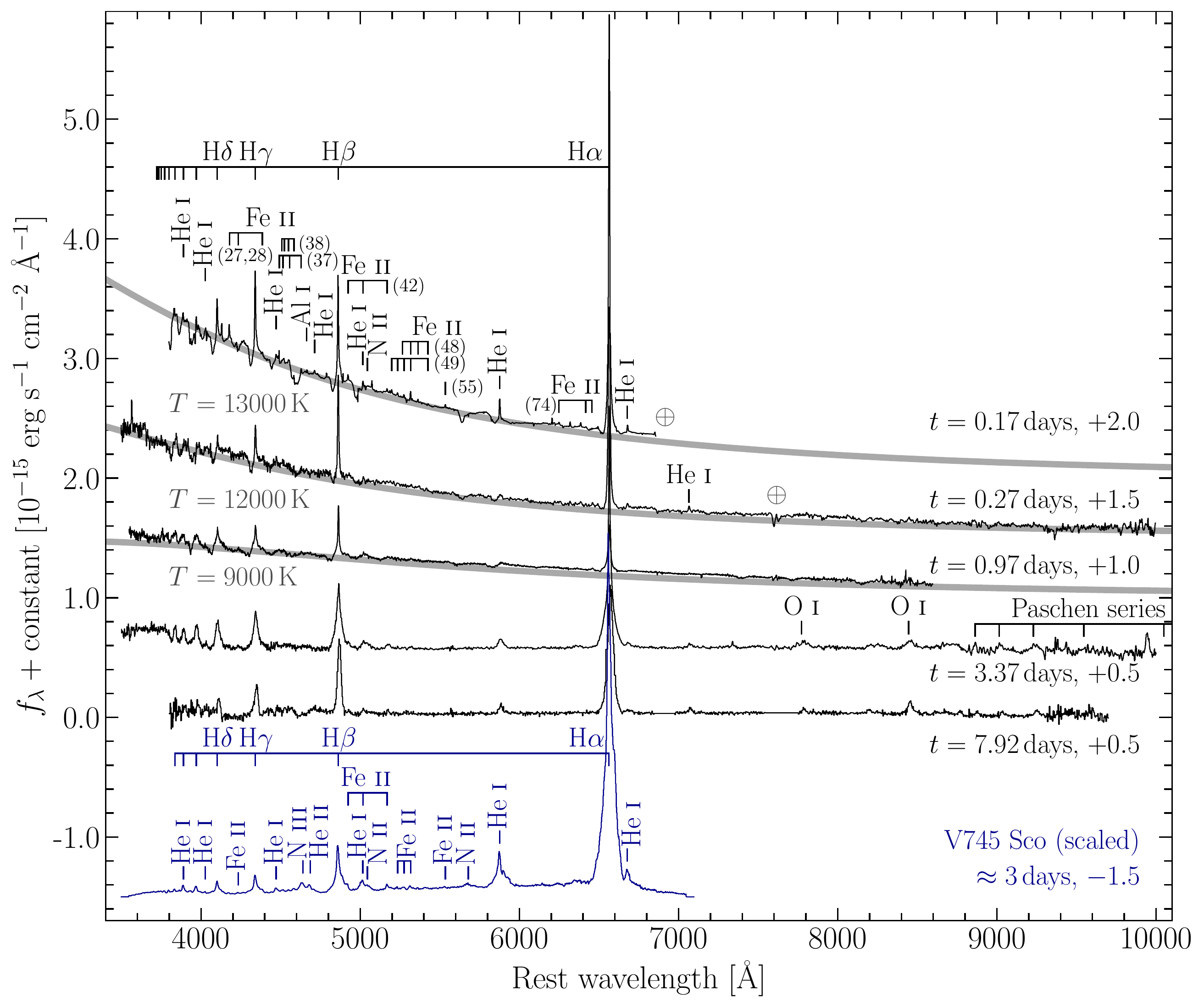}
\caption{\label{fig:spec}
Optical spectroscopic sequence of AT\,2019qyl. The spectra have been dereddened for Galactic extinction with a \citet{fitzpatrick99} extinction law with $R_V = 3.1$ and assuming $E(B-V) = 0.01$. The phase of each spectrum and vertical offset in flux (for clarity) are indicated along the right side of the figure. Prominent emission features identified primarily from the line lists of \citet{williams12} are labeled. Blackbodies that approximate the continuum emission at $t=0.17$, $0.27$, and $0.97$\,days with $T = 13000$, $12000$, and $9000$\,K, respectively, are shown as the thick gray curves. For comparison, we show the 1979 August 1 optical spectrum of V745~Sco, a recurrent nova in a symbiotic binary system, taken $\approx$3\,days after maximum from \citet{williams91}. Prominent features in this spectrum are also labeled, and we note a strong similarity to the similar phase, postmaximum spectrum of AT\,2019qyl from $t=3.37$\,days.
}
\end{figure*}

We obtained a sequence of five optical spectra between 2019 September 26.29 and 2019 October 4.04, spanning from $1.9$\,hr to 8\,days after discovery. Our earliest spectrum was obtained with the Gemini Multi-Object Spectrographs (GMOS; \citealp{hook04,gimeno16}) on the 8.1\,m Gemini South Telescope as part of our program for rapid ToO observations of newly discovered transients with DLT40 (PID GS-2019B-Q-125; PI D.\ Sand). In addition, we obtained two spectra with the Las Cumbres Observatory FLOYDS spectrograph on the 2\,m Faulkes Telescope North (FTN) on Haleakala in Hawaii and one spectrum with the Robert Stobie Spectrograph (RSS; \citealp{burgh03,kobulnicky03}) on the 10\,m Southern African Large Telescope (SALT; \citealp{buckley06}). Our last optical spectrum was obtained with the Alhambra Faint Object Spectrograph and Camera (ALFOSC) on the 2.56\,m Nordic Optical Telescope (NOT) at the Spanish Observatorio del Roque de los Muchachos on La Palma using the  low-resolution Gr4 300\,lines\,mm$^{-1}$ grism (OPTICON 19B-053; PI N.\ Blagorodnova). The spectra were reduced using standard techniques including wavelength calibration with arc-lamp spectra and flux calibration using spectrophotometric standard stars. In particular, (1) we used standard tasks in the Gemini IRAF package\footnote{\url{http://www.gemini.edu/sciops/data-and-results/processing-software}} for the GMOS spectrum following procedures provided in the GMOS Data Reduction Cookbook;\footnote{\url{http://ast.noao.edu/sites/default/files/GMOS_Cookbook/}} (2) we followed the procedures of \cite{valenti14} for the FLOYDS spectra, and (3) used the Python package \textsc{PypeIt}\footnote{\url{https://pypeit.readthedocs.io/en/latest/}} for the ALFOSC spectrum \citep{prochaska20,pypeit_zenodo}. Our optical spectra are shown in Figure~\ref{fig:spec}.

\begin{figure}
\centering
\includegraphics[width=0.49\textwidth]{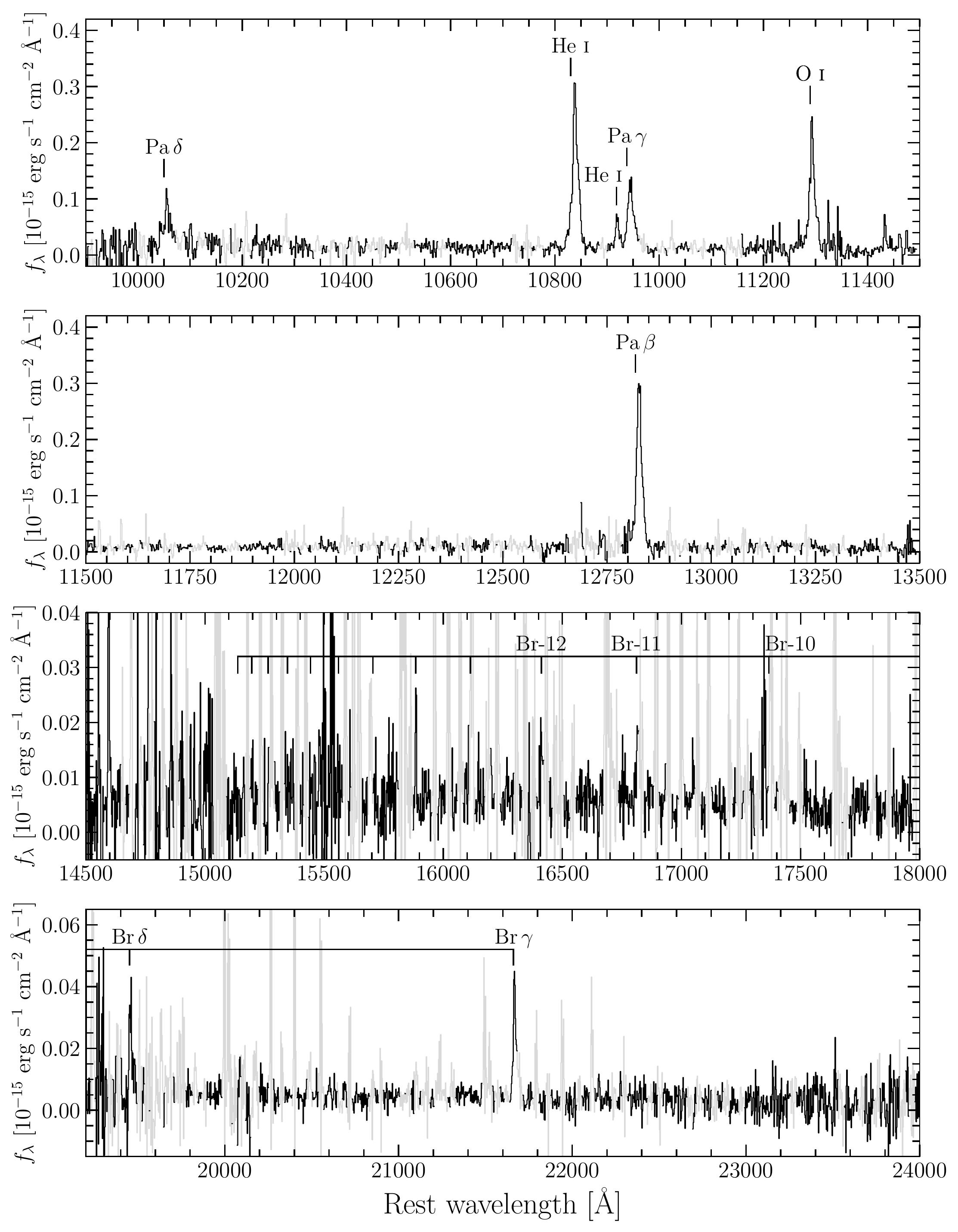}
\caption{\label{fig:NIR_spec}
The late-time NIR spectrum of AT\,2019qyl from $t=69.14$\,days. We label prominent emission features of \ion{H}{1}, \ion{He}{1} ($\lambda 10830$, $\lambda 10913$), the Ly$\beta$ fluoresced \ion{O}{1} ($\lambda 11287$). Spectral bins of low S/N due to contamination by bright OH airglow emission lines are plotted in light gray.
}
\end{figure}

A late-time NIR spectrum was obtained with the Near-Infrared Echellette Spectrometer\footnote{\url{https://www2.keck.hawaii.edu/inst/nires/}} (NIRES) on the 10\,m Keck 2 Telescope on Maunakea in Hawaii on 2019 December 4.25, $\approx$69\,days postdiscovery. NIRES uses a $0\farcs55$ slit and provides wavelength coverage from 9500 to 24600\,\AA\ across five spectral orders at a mean resolution of $R = 2700$. During the observations, we nodded the target along the slit between exposures in a standard ABBA pattern to allow for accurate subtraction of the sky background. Observations of the A0\,V telluric standard star HIP~4064 near the target position were also taken immediately preceding the science target observation for flux calibration and correction of the strong NIR telluric absorption features. The data were reduced, including flat fielding, wavelength calibration, background subtraction, and 1D spectral extractions steps, using a version of the IDL-based data reduction package Spextool developed by \citet{cushing04}, updated by M.\ Cushing specifically for NIRES. Telluric corrections and flux calibrations were performed with the standard-star observations using the method developed by \citet{vacca03} implemented with the IDL tools \textsc{xtellcor} or \textsc{xtellcor\_general} developed by \citet{cushing04} as part of Spextool. The NIR spectrum is shown in Figure~\ref{fig:NIR_spec}.

A summary of all our spectroscopic observations is provided in Table~\ref{table:spec}.

\begin{deluxetable*}{lcrlccc}
\tablecaption{Log of Spectroscopic Observations \label{table:spec}}
\tablehead{\colhead{UT Date} & \colhead{MJD} & \colhead{Phase} & \colhead{Tel./Instr.} & \colhead{Range} & \colhead{Resolution} & \colhead{H$\alpha$ Resolution} \\ 
\colhead{} & \colhead{} & \colhead{(days)} & \colhead{} & \colhead{(\AA)} & \colhead{($\lambda / \delta \lambda$)} & \colhead{(km\,s$^{-1}$)}}
\startdata
2019 Sep 26.29 & 58752.29 & 0.17  & Gemini S/GMOS     & 3750--7000   & 560       & 540 \\ 
2019 Sep 26.39 & 58752.39 & 0.27  & FTN/FLOYDS        & 3200--10000  & 240--420  & 900 \\ 
2019 Sep 27.09 & 58753.09 & 0.97  & SALT/RSS          & 3200--9000   & 500--1700 & 210 \\ 
2019 Sep 29.49 & 58755.49 & 3.37  & FTN/FLOYDS        & 3200--10000  & 240--420  & 900 \\ 
2019 Oct 04.04 & 58760.04 & 7.92  & NOT/ALFOSC        & 3200--9600   & 360       & 830 \\ 
2019 Dec 04.25 & 58821.25 & 69.13 & Keck 2/NIRES      & 9500–24600   & 2700      & \nodata \\ 
\enddata
\end{deluxetable*}

\section{The Transient}\label{sec:transient}
This section presents our analysis of the host environment of the transient and its posteruption photometric and spectroscopic evolution. 

\subsection{Host Extinction and Environment}\label{sec:host}
The transient is located in a spiral arm of NGC\,300, as shown in Figure~\ref{fig:pre-imaging}, but it does not appear to be associated with a \ion{H}{2} region or a region of particularly dense star formation. Furthermore, the galaxy is largely face on, and because there is no evidence of a prominent dust lane, any extinction from the host environment will be small. 

Additional evidence from our analysis of the transient points to low or negligible host extinction. \citet{shafter09} suggested a mean color of Galactic novae at maximum light of $B-V = 0.23 \pm 0.06$\,mag from \citet{vandenbergh87}. Correcting for Galactic extinction only, we find $B-V = 0.23 \pm 0.03$\,mag at the time of the light-curve peak (see Section~\ref{sec:lcs} and Table~\ref{table:lc_properties}), implying a low value for the host contribution to the extinction. Additionally, our optical spectra, discussed in Section~\ref{sec:spectral_evolution}, show no evidence for \ion{Na}{1}~D absorption or absorption by the diffuse interstellar bands at the host redshift, which are known to correlate with dust extinction \citep[e.g.,][]{merrill38,hobbs74,phillips13}. Finally, in our analysis in Section~\ref{sec:pre-SED}, we find that the spectral energy distribution (SED) of the quiescent precursor source can be well modeled without any additional extinction from the host. Therefore, we assume a very low or negligible contribution from the host to the total extinction to AT\,2019qyl, and correct only for Milky Way extinction throughout this work. 

We can estimate the metallicity in the environment of AT\,2019qyl using observed metallicity gradients for NGC\,300. For the galactic orientation model of \citet{bresolin09}, the galactocentric distance of AT\,2019qyl is $\approx$3$\farcm$9 ($\approx$2.1\,kpc at the assumed distance of NGC\,300). Using their distribution of oxygen abundances for \ion{H}{2} regions in NGC\,300, we estimate a subsolar metallicity at the location of AT\,2019qyl of $12 + \log(\mathrm{O}/\mathrm{H}) \approx 8.4$, approximately that of the LMC. Using the measurements of \citet{gazak15} based on spectral modeling of red supergiant stars in NGC\,300, we again find a subsolar metallicity of $\log(Z/Z_{\odot}) \approx -0.2$.

\subsection{Time of Eruption}\label{sec:t0}
As described in Section~\ref{sec:disc}, prediscovery nondetections by DLT40 and ATLAS constrain the time of the eruption to $<$0.75\,days before the first detection. The early rise of AT\,2019qyl was tracked at high cadence by DLT40, with 15 measurements made within the first 4.3\,hr of discovery, and is shown in detail in the inset in the lower-left corner of Figure~\ref{fig:lcs}. The source rises from the discovery magnitude of $17.4$ to $17.0$ in 0.89\,days. To better constrain the time of the eruption, we model the early part of the DLT40 light curve (up to the peak) as a power law in flux $f_{\nu} \propto t^{\alpha}$. The best-fitting power-law index to the early rise is $\alpha = 0.16$, which sets the time of eruption, $t_0$, to 2019 September 26.12 (MJD 58752.12), just 2.1\,hr before the first DLT40 detection. \added{We note briefly that the relatively flat rise of the unfiltered DLT40 light curve is driven primarily by the shift of the SED from the UV into the optical and is not reflective of a rise in the bolometric luminosity, which likely occurred on a much faster timescale (see Sections~\ref{sec:lc_rise} and \ref{sec:SED})}. Through this work, we use $t_0$ as our reference epoch for the phase of the transient ($t$). 

\subsection{Posteruption Photometric Evolution}\label{sec:lcs}
As shown in Figure~\ref{fig:lcs}, our comprehensive follow-up observations of AT\,2019qyl with several ground- and space-based facilities (described in Section~\ref{sec:im_obs}) track the multiwavelength photometric evolution in the UV, optical, and IR for the first $\approx$120\,days. Following the peak, the light curves display a smooth, monotonic decline until $t\approx71$\,days, after which a steep dropoff is observed. In similar fashion to the analysis of the recent eruptions of the M31 recurrent nova M31N~2008-12a by \citet{darnley16}, the evolution of AT\,2019qyl can be divided into several distinct phases: the early rise; the fast, initial decline; the slower decline  or ``plateau'' phase; and the subsequent dropoff. Here, we describe each of these phases in detail. 

\subsubsection{The Early Rise to Peak: $t\lesssim 1$\,day}\label{sec:lc_rise}
In the ground-based $g$-band light curve, the observed peak magnitude of the transient is $g = 17.05 \pm 0.01$\,mag (AB system, $M_g = -9.2$\,mag) at $t=0.61$\,days. The $ugri$ data together display a conspicuous trend, with the observed light-curve peaks occurring earlier in the bluer bands ($t = 0.13$\,days in $u$) and later in the redder bands ($t = 0.98$\,days in $i$). A similar trend is apparent in the $U$, $B$, and $V$ bands, successively peaking at $t = 0.25$, $0.73$, and $0.96$\,days. Despite our very high-cadence imaging in the first hours of the eruption, it is still difficult to precisely determine the timing of the peaks owing to the extremely fast early evolution of the transient. Possible differences in calibration of the ground-based $UBV$ and corresponding \textit{Swift} light curves also add uncertainty in estimating the peak times for those bands. Regardless of the above uncertainties, the general trend of a faster rise to peak for the bluer bands, with all bands peaking at times $\lesssim 1$\,day appears robust. This is similar to the rise times of CNe of similar speed classes (see Section~\ref{sec:lc_decline} below), which tend to be $\lesssim 1$\,day \citep{hounsell10,hounsell16}, and especially to those of the RNe M31N~2008-12a and RS~Oph \citep{hounsell10,darnley16}.

This trend is further borne out in the UV: in the \textit{Swift}/UVOT $uvw1$ filter, we find the transient to be in a steep decline between $t=0.25$ and $0.71$\,days, fading by $0.9$\,mag from the observed peak at $17.42 \pm 0.08$ mag. Comparing the ground-based $u$ and \textit{Swift} $U$ light curves, the UV peak may have occurred near the time of our earliest \textit{Swift} observations between $t = 0.13$ and $0.25$\,days. \citet{darnley16} noted a correlation between times to peak and wavelength for M31N~2008-12a, with shorter times to peak for the bluer bands. Despite similar rise time in the visible, we find the UV light curves of AT\,2019qyl peaked even faster compared to a rise time of $\approx$0.5--0.7\,days for M31N~2008-12a in $uvw1$. This indicates significant early evolution of the SED of the transient (see Section~\ref{sec:SED}), and a significant decrease in the temperature of the photosphere. 

This is mirrored in the evolution of our optical spectra, which show an early hot, blue continuum that quickly fades within the first day after eruption (see Section~\ref{sec:spectral_evolution}). 

\subsubsection{The Initial Decline: Days $\approx$1--10}\label{sec:lc_decline}
Following the peaks in the optical bands by $t=0.94$\,days, we observe a rapid initial decline until a distinct break is observed in the optical $BVgr$ bands between $\approx$4 and 11\,days. We fit the $BVgri$ light curves for phases of 1--80\,days with broken power laws in flux, allowing the time of the break, $t_{\mathrm{break}}$, to be determined by the fit. The results of these fits are shown in Figure~\ref{fig:lcs}, and the best-fitting power-law indices to the early and late portions of the decline are given in Table~\ref{table:lc_properties} as $\alpha_1$ and $\alpha_2$, respectively. The $BVg$-band light curves all behave similarly, showing an initial steep decline ($\alpha_1 \approx -0.9$ to $-1.1$) and light-curve breaks between $t_{\mathrm{break}} \approx 5$--$11$\,days. We note that the break in the $r$-band light curve occurs first at $t=4.2$\,days. Following this, it declines much more slowly than the others, likely as a result of strong H$\alpha$ emission that dominates the optical spectra (see Section~\ref{sec:spectral_evolution}). The break in the $i$-band light curve occurs later at $t_{\mathrm{break}} = 30.0$\,days, after which it tracks with the more slowly declining $r$ band, possibly indicating that strong emission lines may also begin to dominate the $i$-band flux at this phase. As our $u$- and $U$-band coverage lasts only until $t = 11$\,days, we fit the light curves with only a single power law and find similar decline rates to the $BVg$ bands during the initial decline. 

The timescale of the optical light-curve decline for novae are traditionally described by the time for the light curve to fall by 1, 2, and 3\,mag from the peak, which we refer to as $t_1$, $t_2$, and $t_3$, respectively. While this simple parameterization does not capture the full diversity of nova light curves \citep[see, e.g.,][]{strope10}, it is useful for placing AT\,2019qyl in the context of known Galactic novae as well as the growing sample of extragalactic events. We measure these using linear interpolations of the photometric measurements that bracket the 1, 2, and 3\,mag drops from the peak and list our results in Table~\ref{table:lc_properties}. In the $V$ band, we find $t_1$, $t_2$, and $t_3$ to be $1.0$, $3.5$, and $10.3$\,days. This would classify AT\,2019qyl as a ``very fast'' nova ($t_2 < 10$\,days; \citealp{payne-gaposchkin57}). As discussed below in Section~\ref{sec:RNe}, this places AT\,2019qyl among the fastest known novae, which are generally believed to occur on massive WDs $\gtrsim$1.2~$M_{\odot}$ \citep[e.g.,][]{yaron05}. 

\begin{deluxetable*}{ccccrrrrcc}
\tablecaption{Properties Derived from Optical Light Curves \label{table:lc_properties}}
\tablehead{\colhead{Filter} & \colhead{$t_\mathrm{peak}$} & \colhead{$m_\mathrm{peak}$\tablenotemark{a,b}} & \colhead{$M_\mathrm{peak}$\tablenotemark{a,b}} & \colhead{$t_1$\tablenotemark{c}} & \colhead{$t_2$\tablenotemark{c}} & \colhead{$t_3$\tablenotemark{c}} & $t_{\mathrm{break}}$ & \colhead{$\alpha_1$\tablenotemark{d}} & \colhead{$\alpha_2$\tablenotemark{d}} \\ 
\colhead{} & \colhead{(days)} & \colhead{(mag)} & \colhead{(mag)} & \colhead{(days)} & \colhead{(days)} & \colhead{(days)} & \colhead{(days)} & \colhead{} & \colhead{} }
\startdata
$U$  & 0.25 & 16.43 (0.08) & $-9.86$ & 1.3  &  3.9 & 11.6    & \nodata & $-1.01$ & \nodata \\
$B$  & 0.73 & 17.35 (0.09) & $-8.94$ & 1.2  &  4.0 & 17.4    & 5.2     & $-1.10$ & $-0.74$ \\
$V$  & 0.96 & 17.07 (0.02) & $-9.22$ & 1.0  &  3.5 & 10.3    & 6.5     & $-1.14$ & $-0.75$ \\
$u$  & 0.13 & 17.34 (0.03) & $-8.95$ & 2.2  &  6.1 & \nodata & \nodata & $-0.92$ & \nodata \\
$g$  & 0.61 & 17.05 (0.01) & $-9.24$ & 1.2  &  3.4 & 14.23   & 11.2    & $-1.03$ & $-0.66$ \\
$r$  & 0.62 & 17.09 (0.01) & $-9.20$ & 2.7  & 38.4 & \nodata & 4.2     & $-0.73$ & $-0.36$ \\
$i$  & 0.98 & 17.12 (0.02) & $-9.17$ & 2.7  &  8.9 & 23.5    & 30.0    & $-0.92$ & $-0.40$ \\
\enddata
\tablenotetext{a}{Vega magnitudes given for $UBV$ and AB magnitudes given for $ugri$. $1\sigma$ uncertainties given in parentheses.}
\tablenotetext{b}{Corrected for Galactic extinction toward NGC\,300 assuming $E(B-V) = 0.01$.}
\tablenotetext{c}{$t_1$, $t_2$, and $t_3$ are defined as the times for the light curves to decline by 1, 2, and 3 mag, respectively, postpeak.}
\tablenotetext{d}{Best-fitting power-law index to flux measurements between $1 \leq t~[\mathrm{days}] \leq t_{\mathrm{break}}$ for $\alpha_1$ and $t_{\mathrm{break}} \leq t~[\mathrm{days}] \leq 80$ for $\alpha_2$.}
\end{deluxetable*}

\subsubsection{The Slow-decline Phase and Dropoff: $t \gtrsim 10$\,days}\label{sec:lc_late}
Following the steep, initial decline, AT\,2019qyl enters a phase of slower decline after $t \approx 10$\,days, as indicated by the shallower power-law indices found in this phase for the $BVgri$ light curves ($\alpha_2$ in Table~\ref{table:lc_properties}). Beyond the optical bands, a similar change in the decline rate is also apparent in the IR [3.6] and [4.5] bands from \textit{Spitzer}/IRAC. This phase is sometimes referred to as a ``plateau'' as defined by \citet{strope10} for class ``P'' nova light curves, though it includes events like AT\,2019qyl, which continue to decline through the plateau phase, though at a slower rate. This phase lasts until at least $t \approx 71$\,days in the optical light curves. Subsequently, the $r$-band enters a phase of steeper decline between $t = 94$\,days and the end of our ground-based photometric monitoring at $127$\,days. Similarly, the late-time WFC3/UVIS F555W observation with \textit{HST}, in comparison with the last ground-based $V$-band measurement, also indicates a steep dropoff occurred by $t=122$\,days. This suggests the drop in observed flux is not only driven by a drop in H$\alpha$ luminosity, but likely reflects a drop in the optical continuum emission. 

\citet{hachisu06a,hachisu07} have suggested a ``universal decline law'' for novae under the assumption that free-free emission from the optically thin, expanding nova shells dominates the continuum flux. Their model consists of a broken power law for the mid- and late-time light curve, with the time of the break corresponding to a drop in the wind mass-loss rate at the end of steady hydrogen-burning on the WD surface. The time of the break in their model depends primarily on the WD mass (and weakly on composition), providing an observationally convenient method to estimate WD masses for well-observed novae. In this model, the initial transition to the slow optical decline and subsequent dropoff are timed with the emergence and turnoff of super-soft X-ray emission powered by the continued nuclear burning, also used as a proxy for the WD mass \citep[e.g.][]{henze11,schwarz11,wolf13}. We show their predicted power laws in comparison to our light curves in Figure~\ref{fig:lcs}, and, while the measured power-law indices for $t > 10$\,days ($\alpha_2$ in Table~\ref{table:lc_properties}) do not precisely match the predicted value, we note a qualitative similarity between the predictions and the observed slow-decline phase and subsequent dropoff. 

This light-curve evolution is very similar to the Galactic RN RS~Oph. During its 2006 eruption, its light curve transitioned from an initial steep decline into a slow decline or plateau, lasting until day 83, and subsequently a steep dropoff \citep[e.g.,][ and references therein]{schaefer10a}. In the context of other very fast-evolving and Galactic RG novae for which WD masses have been estimated via this method (e.g., $\approx$1.35\,$M_{\odot}$ for RS Oph; \citealp{hachisu06b,hachisu18}; see Section~\ref{sec:RNe} for further discussion), we may infer a similarly massive WD for AT\,2019qyl. We note, however, that the applicability of this simple model is limited by several factors that can substantially affect the broadband light curves, such as the interaction of the nova ejecta with the preexisting wind of the RG companion, internal shocks (Sections~\ref{sec:line_profiles} and \ref{sec:shocks}), and other complicating factors such as dust formation or the contribution of strong emission lines.

The decline in the IR [3.6] and [4.5] light curves within the first 67 days appears consistent with the one observed in the optical bands. We note here that our \textit{Spitzer} photometry of the active transient is based on reference-subtracted images, and hence the contribution of the underlying RG (Section~\ref{sec:IR_SED}) companion has been removed. Some slow novae on CO WDs, particularly those of the DQ Her class, are characterized by the formation of optically thick dust shells in the ejecta. Notable examples are NQ Vul \citep{ney78}, LW Ser \citep{gehrz80a}, and V5668 Sgr \citep{gehrz18}. These shells are observed to reradiate the maximum luminosity of the nova as IR emission at $\approx$50--80\,days after the eruption. Faster CO novae like V1668 Cyg \citep{gehrz80b} have been observed to produce less dust in optically thin shells, but are still characterized by steep IR brightenings after the onset of dust formation. We do not observe a strong IR brightening that would clearly indicate the formation of dust in AT\,2019qyl for at least the first $\approx$70\,days. As discussed below in Section~\ref{sec:spectral_evolution}, the NIR spectrum taken at a similar phase of $t=69.14$\,days does not show evidence for a red continuum beyond $\approx$2\,$\mu$m that would indicate thermal emission from newly formed dust. 

\subsection{SED Evolution}\label{sec:SED}
\begin{figure}
\centering
\includegraphics[width=0.49\textwidth]{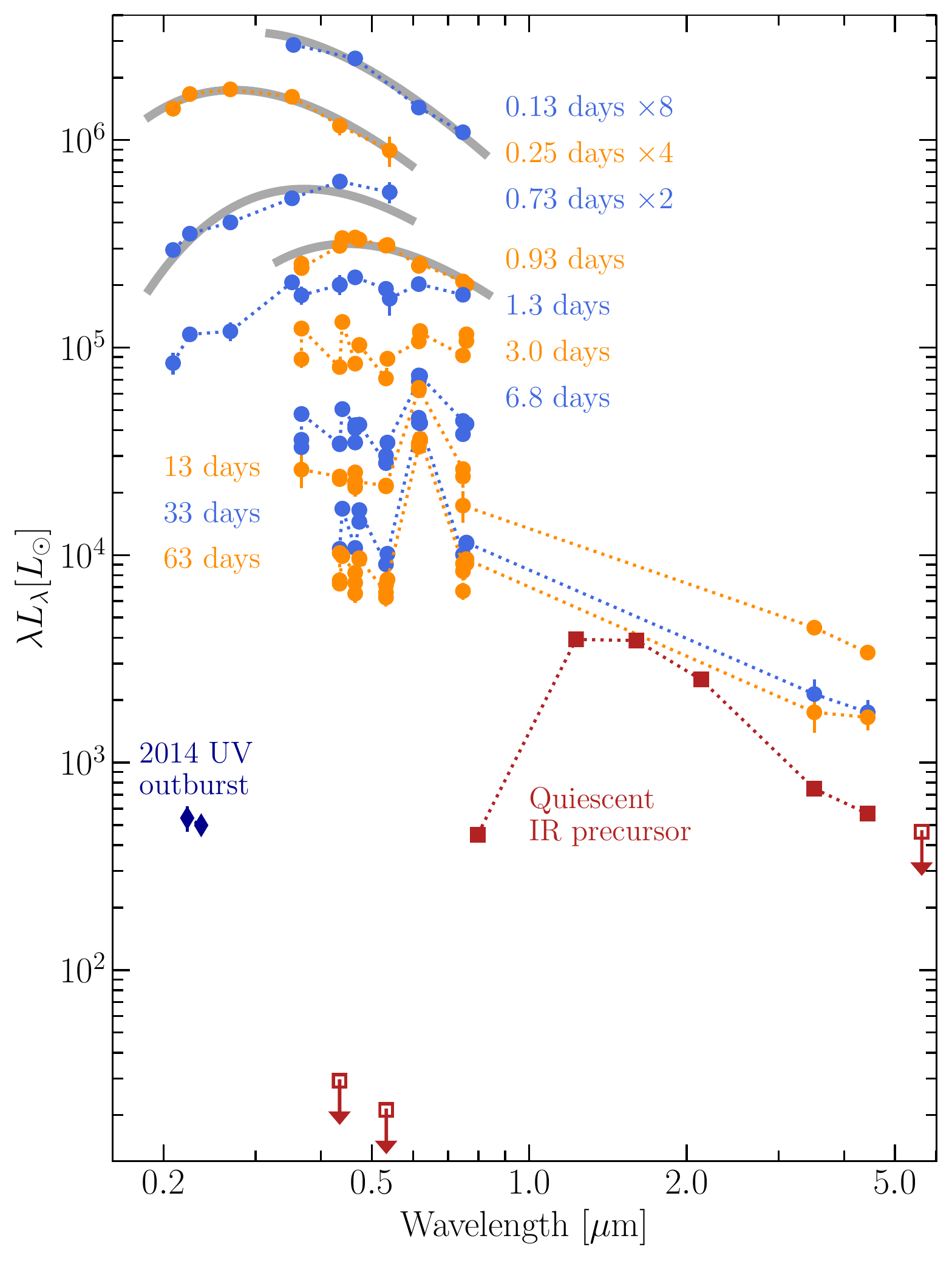}
\caption{\label{fig:SED_evolution}
The SED evolution of AT\,2019qyl from photometry is shown for several representative phases from $t=0.13$ to $63$\,days in alternating colors for clarity (blue and orange circles). SEDs for the earliest three epochs have been scaled up by the factors indicated on the figure to avoid crowding. Blackbody approximations to the data are shown as the thick gray curves for the first four epochs, with fit parameters provided in Table~\ref{table:bb_fits}. For comparison, we also show the SEDs of the quiescent IR precursor source (dark red squares) and the prior 2014 UV outburst (dark blue diamonds; see Section~\ref{sec:pre-SED} and Figure~\ref{fig:pre_SED} for details).
}
\end{figure}

We constructed quasi-contemporaneous SEDs from our photometry at several representative epochs in the evolution of AT\,2019qyl between $t=0.13$ and $63$\,days. We consider observations in different photometric bands to be contemporaneous if the difference in time between them is less than $1/10$th the age of the transient at that phase and adopt the average phase of the included measurements at each epoch. We converted our extinction-corrected photometric measurements to band luminosities ($\lambda L_{\lambda}$) at the assumed distance to NGC\,300 and for the appropriate zero-magnitude fluxes and nominal effective wavelengths for each filter. The resulting SEDs are shown in Figure~\ref{fig:SED_evolution}. 

At the earliest phases ($t\lesssim0.3$~days), the SED peaks in the UV below $\lesssim 0.3$\,$\mu$m. Following this, the peak of the SED quickly shifts into the optical by $t = 0.93$\,days. This is reflected in the rapid decline of the UV light curves as the optical light curves rise to maximum light as described above (Section~\ref{sec:lc_rise}). We fit blackbody approximations to these early SEDs using a custom Markov Chain Monte Carlo (MCMC) code \texttt{BBFit}\footnote{\url{https://github.com/nblago/utils/blob/master/src/model/BBFit.py}} based on \texttt{emcee} \citep{foreman-mackey13}. The results for the best-fitting models and 1$\sigma$ (68\%) confidence intervals are given in Table~\ref{table:bb_fits}. \added{We caution that blackbody models offer only a crude approximation to early nova spectra, where line-blanketing effects, particularly in the UV, and departures from local thermodynamic equilibrium can be significant \citep[e.g.,][]{hauschildt94,hauschildt95}. Still, we find the simple blackbody analysis described here to be illustrative.}

Initially, we infer high blackbody temperatures of $T_{\mathrm{BB}} = 11690$ and $13600$\,K at $t=0.13$ and $0.25$\,days, respectively. The apparent rise in temperature between these early epochs may not be reliable owing to the lack of UV data to constrain the fit at $t=0.13$\,days. With a radius of $\approx$10$^{13}$\,cm ($141~R_{\odot}$), we obtain a high bolometric luminosity of $2.4 \times 10^{39}$\,erg~s$^{-1}$ at $t=0.25$\,days, assuming the observed UV-optical flux accounts for the bulk of the radiated emission. This is $\gtrsim$10 times the electron-scattering Eddington luminosity for a massive WD of $L_{\mathrm{Edd}} = 1.8 \times 10^{38} \left(\frac{M}{1.2~M_{\odot}} \right)$~erg~s$^{-1}$. At the time of the optical light-curve maxima at $t=0.93$\,days, we find that the SED can now be approximated by a cooler blackbody with $T_{\mathrm{BB}}\approx 8210$\,K and $R_{\mathrm{BB}}\approx325~R_{\odot}$. We thus infer a photospheric velocity of $\approx$2200\,km\,s$^{-1}$ during the rise to peak. This is similar to that inferred from the velocity of the P Cygni absorption features observed for prominent \ion{H}{1} and \ion{He}{1} emission lines in the early spectra of $\approx$1500--4000\,km\,s$^{-1}$ (see Section~\ref{sec:line_profiles}). The inferred bolometric luminosity decreases somewhat to $\approx$(1.6--$1.7) \times 10^{39}$~erg~s$^{-1}$ but remains substantially super-Eddington. Overall, the early SED evolution of AT\,2019qyl is consistent with the so-called ``fireball'' phase, consisting of an increasing radius and decreasing temperatures as the optically thick nova envelope rapidly expands and cools \citep{gehrz88a, hauschildt94}. 

\begin{deluxetable}{cccccc}\label{table:bb_fits}
\tablecaption{Blackbody Fits to Early SEDs}
\tablehead{\colhead{Phase} & \colhead{$T_{\mathrm{BB}}$} & \multicolumn{2}{c}{$R_{\mathrm{BB}}$} & \multicolumn{2}{c}{$L_{\mathrm{BB}}$} \\
\cmidrule(lr){3-4}\cmidrule(lr){5-6}
\colhead{(days)} & \colhead{(K)} & \colhead{($10^{13}$\,cm)} & \colhead{($R_{\odot}$)} & \colhead{($10^{39}$\,erg~s$^{-1}$)} & \colhead{($10^{5} L_{\odot}$)}}
\startdata
0.13 & $11690^{+70}_{-70}$   & $1.20^{+0.01}_{-0.01}$ & $172^{+1}_{-1}$ & $2.0^{+0.1}_{-0.1}$    & $5.0^{+0.2}_{-0.2}$ \\
0.25 & $13600^{+200}_{-200}$ & $0.98^{+0.03}_{-0.03}$ & $141^{+4}_{-4}$ & $2.4^{+0.3}_{-0.3}$    & $6.1^{+0.8}_{-0.7}$ \\
0.73 & $9900^{+100}_{-100}$  & $1.51^{+0.04}_{-0.04}$ & $217^{+6}_{-6}$ & $1.6^{+0.2}_{-0.2}$    & $4.1^{+0.4}_{-0.4}$ \\
0.94 & $8210^{+300}_{-300}$  & $2.26^{+0.01}_{-0.01}$ & $325^{+2}_{-2}$ & $1.68^{+0.04}_{-0.04}$ & $4.3^{+0.1}_{-0.1}$
\enddata
\end{deluxetable}

During the decline phase ($t \gtrsim 1$\,day), comparing the SEDs to that of a blackbody is no longer appropriate as the nova spectra become increasingly dominated by emission lines. This is especially evident in the prominent $r$-band excess that appears after $t \gtrsim 6.8$\,days owing to the large equivalent width of H$\alpha$.

In the IR, as noted above in Section~\ref{sec:lc_late}, we do not observe the development of a strong IR excess between phases of $t\approx 13$--$70$\,days that would indicate the formation of dust in the ejecta. By $t \gtrsim 60$\,days, the IR flux is only a factor of $\approx$2--3 brighter than the quiescent flux of the IR precursor source, which we expect to begin to dominate the NIR SED of AT\,2019qyl as it continues to fade. 

\subsection{Spectroscopic Evolution}\label{sec:spectral_evolution}
Our optical spectroscopic sequence, shown in Figure~\ref{fig:spec}, provides an unprecedented view of the early time evolution during the rise of a rapid nova outburst, with our earliest spectrum taken only $0.17$\,days ($\approx$4\,hr) after the inferred time of the eruption and three spectra by the time of the visible light-curve peak at $t \approx 1$\,day.
The earliest spectrum is characterized by a hot, blue continuum similar to a $T = 13000$\,K blackbody and prominent emission features of H and \ion{He}{1} with P~Cygni absorption components at $v \approx 2000$\,km\,s$^{-1}$. We also note numerous weaker emission features of neutral and singly ionized species, including \ion{N}{2}, \ion{Fe}{2}, and possibly \ion{Al}{1} based on the line lists of \citet{williams12}. There are a number of additional apparent emission features, similar in strength to the \ion{Fe}{2} lines, for which we have not determined secure identifications. These weak emission features also appear narrow, similar to the instrumental resolution of our GMOS spectrum reported in Table~\ref{table:spec} ($v \lesssim 500$\,km\,s$^{-1}$). These features, along with narrow components of H and He, are expected from the slow-moving wind of the companion star, likely an O-rich asymptotic-giant-branch (AGB) star (see Section~\ref{sec:IR_SED}), flash-ionized by the explosion on the WD.

In analogy with the P Cygni absorption components of the dominant emission lines, we note additional absorption features that may be associated with weaker lines in a high-velocity outflow. For example, the absorption feature near 4600\,\AA\ may be due to blended absorption of the nearby \ion{Fe}{2} (37; $\lambda$ 4629) and \ion{Al}{1} ($\lambda$ 4663), and the absorption feature near 4975\,\AA\ is likely due to a combination of the nearby \ion{Fe}{2} (42; $\lambda$ 5018), \ion{He}{1} ($\lambda$ 5016), and \ion{N}{2} ($\lambda$ 5045). There is also a conspicuous absorption feature near 5650\,\AA\ that does not appear to have an associated emission component. 

As the spectrum evolves through the transient peak around $t = 1.0$\,day, we note a rapid cooling of the continuum emission, from $\approx$12000\,K in the $t=0.27$\,day spectrum to $\approx$9000\,K in the $t=0.97$\,day spectrum. This is similar to the temperature evolution of blackbody approximations to the early SEDs discussed above in Section~\ref{sec:SED}. During this time, the H and \ion{He}{1} P Cygni features display a complex evolution and develop multiple higher-velocity components ($v \approx 2500$--$4500$\,km\,s$^{-1}$), which we examine in more detail below in Section~\ref{sec:line_profiles}. In a similar fashion, the unidentified absorption feature near 5650\,\AA\ (in the rest frame of the host) appears shifted to the blue, and three additional absorption features near 7400, 7715, and 8150\,\AA\ are now apparent. The velocity structure and evolution of these features are further discussed in Section~\ref{sec:unIDd_abs}. 

During the light-curve decline phase at $t \gtrsim 1$\,day, the hot continuum has faded, and the spectra are dominated entirely by emission features. The strong H and He features have also transitioned to pure emission lines with a broader profile with FWHM velocities of $\approx$2000\,km\,s$^{-1}$. We also note the emergence of \ion{O}{1} emission lines ($\lambda \lambda$7773, 8446) by $t = 3.37$\,days. At this phase, the spectrum bears a strong resemblance to that of the 1979 outburst of V745~Sco, a well-studied recurrent nova in a symbiotic binary system \citep{williams91}. Based on this comparison, we note that features of \ion{N}{2} and weaker \ion{Fe}{2} are still present, and identify additional emission features of \ion{He}{2} and \ion{N}{3}. The most prominent emission features at $t=3.37$\,days after H are those of He and N, consistent with an He/N spectral classification. 

In the late-time, NIR spectrum at $t = 69.14$\,days, we detect strong nebular emission features of H, \ion{He}{1}, and \ion{O}{1} with narrower, symmetric profiles. In addition to the strong \ion{He}{1} $\lambda 10830$ emission, we also identify the $\lambda 10913$ $^3D$-$^3F^o$ transition of \ion{He}{1}. We do not observe any NIR lines of \ion{C}{1} that are hallmarks of \ion{Fe}{2}-class novae (strongest lines at 1.166, 1.175, 1.689\,$\mu$m, and several lines between 1.72 and 1.79\,$\mu$m \citealp{das08,banerjee12}), consistent with our optical spectral classification of AT\,2019qyl as an He/N nova. The spectrum is still dominated by the nebular emission of the nova. We do not see the cool continuum and absorption features of the underlying AGB companion, or a thermal continuum indicative of newly formed, warm dust.

\subsubsection{Emission-line Profile Evolution}\label{sec:line_profiles}
\begin{figure*}
\centering
\includegraphics[width=\textwidth]{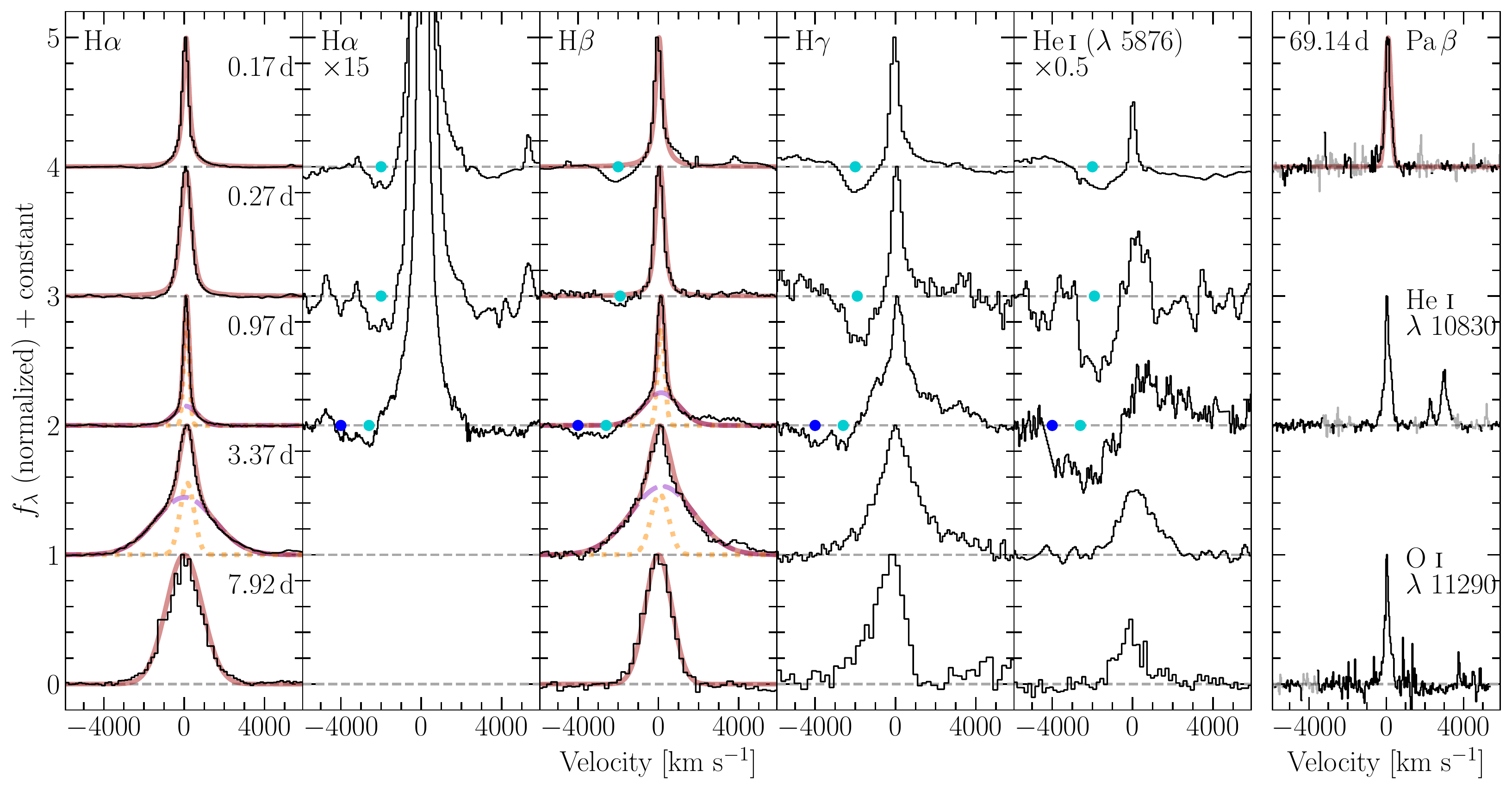}
\caption{\label{fig:line_profiles}
In the left grid of panels, we show the evolution of the H$\alpha$ (with a magnified view in the panel second from the left), H$\beta$, H$\gamma$, and \ion{He}{1} ($\lambda$5876) line profiles as labeled along the top of the figure. The line profiles (black curves) have been continuum subtracted, normalized, and offset vertically by constant values for clarity. Time increases from top to bottom in each of these five panels as labeled in the leftmost one. In the panel offset to the right, we show the $t=69.14$\,day profiles of the near-IR Pa$\beta$, \ion{He}{1} ($\lambda$10830), and \ion{O}{1} ($\lambda$11290) lines offset vertically for clarity. Best-fit models to the H$\alpha$, H$\beta$, and Pa$\beta$ profiles (see main text) are overplotted with the data as thicker red curves. Individual intermediate-width (dashed purple curves) and narrow (dotted orange curves) emission components of the fits are also shown for the $t=0.97$ and $3.37$\,day profiles. Finally, we highlight the velocities of the slow (cyan circles) and fast (dark blue circles) ejecta inferred from the observed P Cygni absorption features in the earliest three spectra.  
}
\end{figure*}

\begin{figure}
\centering
\includegraphics[width=0.5\textwidth]{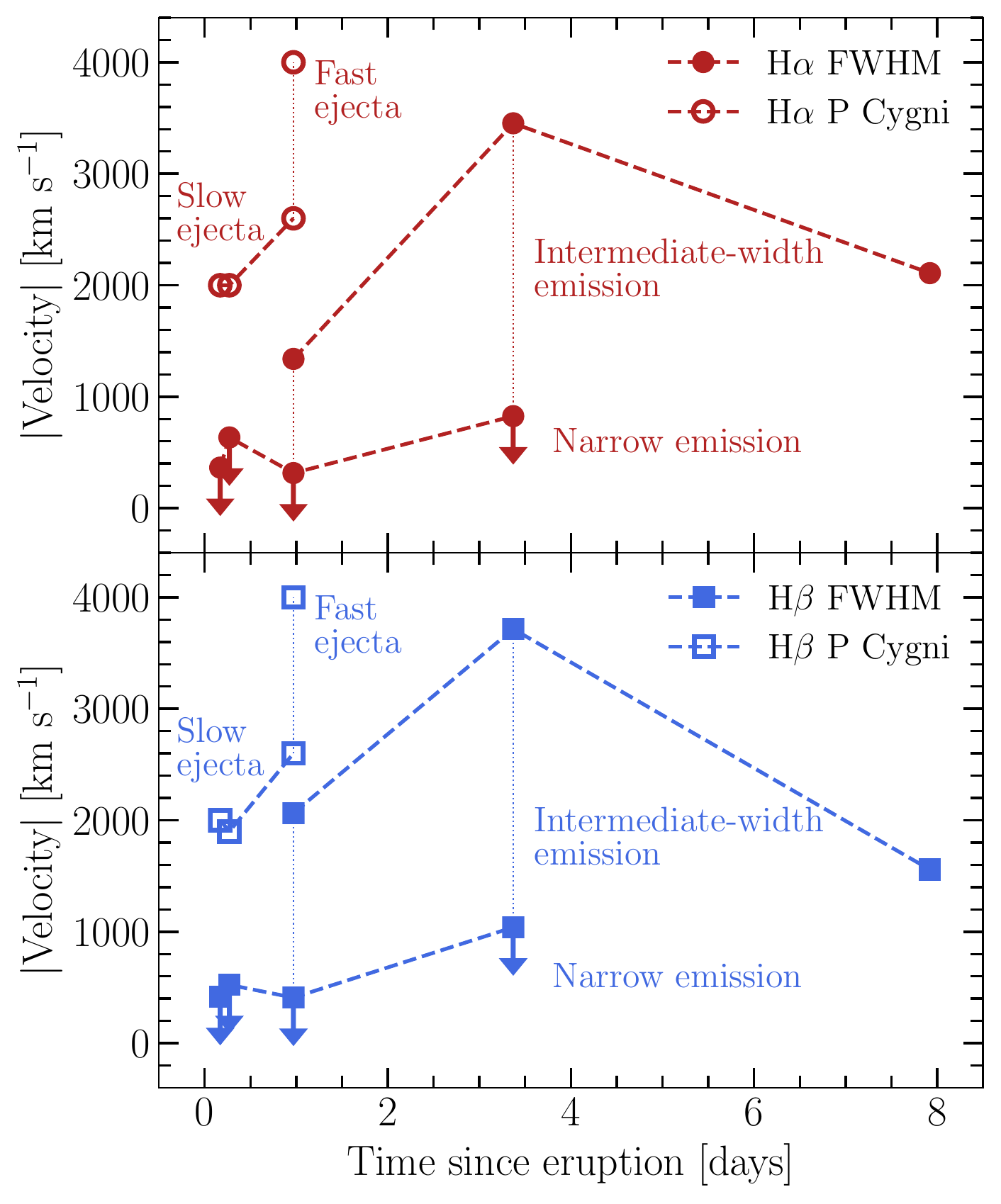}
\caption{\label{fig:line_vel}
Velocity evolution of the components of the H$\alpha$ (top panel, red circles) and H$\beta$ (bottom panel, blue squares) line profiles. FWHM velocities of individual narrow and intermediate-width emission components are shown as filled symbols, where velocities comparable to the instrumental resolution are indicated as upper limits with downward arrows. Absolute values of P Cygni absorption component velocities are shown as unfilled symbols. 
}
\end{figure}

Our very early spectroscopic observations reveal a complex velocity structure and evolution in the strong H and \ion{He}{1} lines (see Figure~\ref{fig:line_profiles}). We examine the line profile evolution in detail here, and our velocity measurements of individual emission and absorption components of H$\alpha$ and H$\beta$ are further illustrated in Figure~\ref{fig:line_vel}. Of particular interest is the evolution of the P Cygni absorption components within the first $t \lesssim 1$\,day, which probe the velocities of the expanding outflows of gas launched by the eruption. In the earliest two spectra at $t=0.17$ and $t=0.27$ days, each of the profiles shows a narrow emission core consistent with the rest-frame zero velocity of NGC\,300. The emission profiles in the first spectrum are approximately Lorentzian with $\mathrm{FWHM} = 360$~km~s$^{-1}$ (see fits for H$\alpha$ and H$\beta$ in Figure~\ref{fig:line_profiles}). The wings of the profiles appear to extend to $\approx$2000\,km\,s$^{-1}$. In the second spectrum, we require broader Voigt profiles with $\mathrm{FWHM} \approx 600$~km~s$^{-1}$ to fit the lines, which we attribute to the lower instrumental resolution. The \added{unresolved,} narrow emission-line cores likely arise from recombination in the slow-moving, ionized wind of the companion. The broader wings seen in the $t=0.17$-day spectrum may be the result of interaction as the nova shock wave propagates into the ambient medium of the companion wind. This is analogous to the early spectra of Type IIn SNe that show signatures of interaction with dense circumstellar material \citep[CSM; e.g.,][]{schlegel90,filippenko97} and is consistent with Galactic examples of novae in symbiotic systems like RS Oph \citep[e.g.][]{bode87,evans08} and V407 Cyg \citep{munari11}. The lines also show a blueshifted P Cygni absorption component at $\approx$2000\,km\,s$^{-1}$ \added{indicative of the expanding nova ejecta}, where the deepest portion of the absorption component appears to be spread over a range of velocities between $\approx$1500 and 2500\,km\,s$^{-1}$. 

Near the time of the optical light-curve peaks at $t=0.97$\,days, we note multiple changes in the profile of the lines. First, while the profiles retain their narrow, \added{unresolved,} emission peak from the ionized companion wind, the base of the emission component has broadened. The emission profiles can now be well approximated by two separate Gaussian components, a narrow and intermediate-width component having FWHM $v \approx 300$--$400$\,km\,s$^{-1}$ and $v \approx 1300$--$2000$\,km\,s$^{-1}$, respectively. The intermediate-width component is similar in velocity to the early P Cygni absorption component seen at $t = 0.17$--$0.27$\,days. \replaced{Now, the P Cygni absorption has shifted to higher velocities between $v\approx2200$\,km\,s$^{-1}$ and $4500$\,km\,s$^{-1}$, and we see clear evidence for distinct velocity components. In particular, we note separate absorption minima indicating a slower component at $\approx$2500\,km\,s$^{-1}$ and fast component at $4000$\,km\,s$^{-1}$.}{Now, the initial $\approx$2000~km~s$^{-1}$ P Cygni absorption component has shifted to $\approx2500$~km~s$^{-1}$, while an additional higher-velocity component at $\approx$4000~km~s$^{-1}$ has also appeared (cyan and blue circles in Figure~\ref{fig:line_profiles}). This is most readily apparent in the relatively clean H$\beta$ line profile.} For any individual line, our choice of where to fit the continuum and potential contamination by other unidentified atomic species would make it difficult to confidently associate these features with real P Cygni absorption components; however, we observe markedly similar profiles in each of the H$\alpha$, H$\beta$, H$\gamma$, and \ion{He}{1} ($\lambda 5876$) lines. This provides strong evidence that these features correspond to real, distinct velocity components in the ejecta. The increase in the slower absorption component from $\approx$2000 to 2500\,km\,s$^{-1}$ within the first day after eruption may correspond to the acceleration of the ejected material or clumps by radiation pressure from the underlying, continued nuclear burning on the WD surface \citep[e.g.,][]{williams08}. \added{It is difficult, however, to imagine a scenario where this acceleration could produce multiple, radially stratified velocity components along the same line of sight from a single, initially homologous ejection. Thus, we argue that the faster 4000~km~s$^{-1}$ component corresponds to a distinct mass ejection that occurred between $t=0.27$ and 0.97\,days.}

By $t=3.37$\,days, the line profiles have transitioned to pure emission features. The lines appear symmetric about their peaks and are notably broader, with wings extending to $\approx$4000\,km\,s$^{-1}$. At this phase, the profiles can again be well approximated by separate \added{Gaussian} components consisting of the \added{unresolved,} narrow core with FWHM $v \approx 360$\,km\,s$^{-1}$ and an even broader base with FWHM $v \approx 3500$\,km\,s$^{-1}$. We suggest that the broadened emission-line profiles arise \replaced{from the shock interaction of physically distinct outflows}{from a combination of the continued interaction of the ejecta with the ambient medium of the companion wind, as well as from the shocked material produced by the collision of the distinct outflows observed as P Cygni absorption components in the earlier spectra}. As a consistency check, we consider a $4000$\,km\,s$^{-1}$ shell launched at $t=1$\,day that would overtake a $2000$\,km\,s$^{-1}$ shell launched at $t_0$ by $t\approx 2$\,days. This is consistent with the observed timing of the appearance of broader, emission-only line profiles, and \replaced{strengthens the interpretation of the higher velocity absorption components in our earlier spectra as indicative of distinct mass ejections occurring during the first $t \approx 1$\,day}{indicates that internal shell--shell collisions are likely to contribute to the production of the broadened emission features at this phase.}

As the nova continues to evolve through the decline phase, the velocities of the broadened emission-line profiles begin to decrease. At $t = 7.92$\,days, the line profiles may now be approximated by a single Gaussian component of FWHM $v \approx 1500$--$2100$\,km\,s$^{-1}$. Finally, in our late-time NIR spectrum, the strongest emission lines of Pa$\beta$, \ion{He}{1} ($\lambda$10830) and \ion{O}{1} ($\lambda$11290) all show similar, approximately Gaussian profiles with FWHM $v \approx 350$\,km\,s$^{-1}$, about a factor of 2 larger than the instrumental resolution of NIRES. This may be interpreted as progressive deceleration of the shock front as it continues to propagate through and sweep up mass from the companion wind. 

\subsubsection{Unidentified Absorption Features}\label{sec:unIDd_abs}
\begin{figure}
\centering
\includegraphics[width=0.5\textwidth]{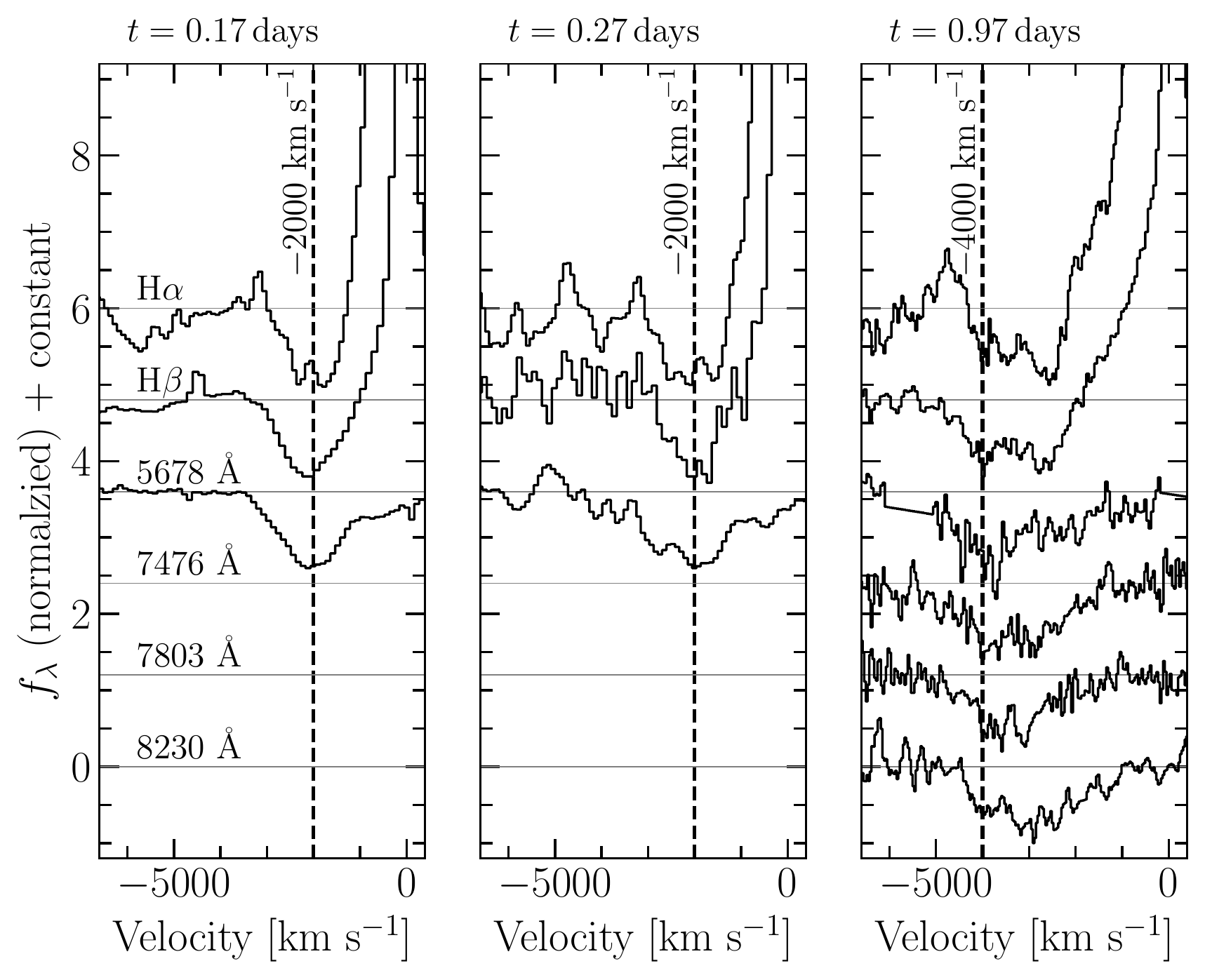}
\caption{\label{fig:unIDd_abs}
Unidentified absorption features are shown in comparison to the P Cygni absorption of H$\alpha$ and H$\beta$ at phases of $t=0.17$, $0.27$, and $0.97$\,days from left to right in each of the three panels. The observed fluxes have been continuum subtracted as in Figure~\ref{fig:line_profiles}, normalized by the depth of the absorption, and shifted by an arbitrary constant for clarity. We infer possible rest wavelengths for each feature (labeled along the left side of the figure) by assuming their velocity evolution matches that of the H features. The dashed vertical lines indicate outflow velocities of $-2000$ (right and center panels) and $-4000$\,km\,s$^{-1}$ characteristic of the inferred multicomponent ejecta. 
}
\end{figure}

As described above, we noted several absorption features in our early-time spectra ($t \lesssim 1$\,day), which do not have corresponding components in emission and for which we have not determined secure identifications. The feature observed near 5650\,\AA\ in the host frame is detected in all three of our early spectra and shows a distinct shift to the blue at $t = 0.97$\,days. As demonstrated in Figure~\ref{fig:unIDd_abs}, this clearly mirrors the evolution of the H$\alpha$ and H$\beta$ P Cygni absorption if we assume a rest ($v=0$) wavelength for the line of $\approx$5678\,\AA. In particular, the shape of the absorption trough is remarkably similar to that of the H$\alpha$ and H$\beta$ profiles at $t=0.17$\,days. Furthermore, for our assumed rest wavelength, it shows a comparable shift in the deepest portion of the absorption from $\approx$2000 to 4000\,km\,s$^{-1}$ and extends across a similar range in velocities ($\approx$1500--4500\,km\,s$^{-1}$) at $t = 0.97$\,days. 

The other three features toward the red end of the optical range are visible in the $t = 0.97$\,day SALT spectrum but were not covered by our earliest GMOS spectrum. They are also not clearly detected in the $t=0.27$\,day FLOYDS spectrum owing to a lower signal-to-noise ratio (S/N). Yet, for assumed rest wavelengths of $\approx$7476, 7803, and 8230\,\AA, these features also appear to have a similar, multicomponent velocity structure to the H$\alpha$, H$\beta$, and $\lambda$5678 feature (Figure~\ref{fig:unIDd_abs}, right panel). While we cannot confirm their velocity evolution owing to a lack of earlier, high-S/N coverage, it is plausible that they arise from the same multicomponent nova ejecta. 

There are some transitions in the line list of \citet{williams12} for CNe that are close to the possible rest wavelengths of these features inferred above. In particular, we note the [\ion{Fe}{4}] $\lambda$5677 and \ion{N}{2} $\lambda$5679 lines for the $\lambda$5678 feature, the \ion{O}{1} $\lambda$7477 line for the $\lambda$7476 feature, and the \ion{O}{1} $\lambda$8227 and \ion{Mg}{2}$\lambda$ 8232 line for the $\lambda$8230 feature.

These features may be akin to the ``transient heavy element absorption'' (THEA) systems, recently recognized to be present in the early phases of CN eruptions---lasting from approximately a few days to weeks after the optical peak---and attributed to elements such as Ti, Ba, Sc, and Y \citep{williams08}. In CNe, these lines typically have comparatively lower velocities ($\approx$400--1000\,km\,s$^{-1}$), and \citet{williams08} and \citet{williams10} suggested they arise from a preexisting circumbinary reservoir of gas, possibly lost from the donor star or accretion disk. Alternatively, as argued in a revised hypothesis by \citet{williams12,williams13}, the THEA features may arise in material stripped from the donor star during the nova eruption. In a recently presented sample of CNe observed spectroscopically before maximum light, \citet{aydi20a} found that the THEA features in one object, V906 Car, tracked the velocity evolution of the slow component of the nova ejecta, supporting the picture that they are associated with the nova flow, rather than preexisting circumbinary gas. They noted, however, that these lines did not display the higher-velocity component associated with the fast ejecta, possibly owing to differences in density or abundances between the ejecta components. These findings differ from the line profiles discussed here, which appear to trace both components of the ejecta. \added{The inferred rest wavelengths of lines seen in AT\,2019qyl do not match those of any of the THEA features reported in these previous works.}

\section{Archival Imaging and the Preeruption Counterpart}\label{sec:arc_im}
The location of AT\,2019qyl has been extensively covered by multiple ground-based and space-based imaging data sets. The available archival \textit{HST} imaging includes ACS/WFC imaging in the F435W, F555W, and F814W filters taken on 2002 July 19 (PID: GO-9492; PI: F.\ Bresolin), and WFC3/UVIS imaging in the F218W and F225W filters taken on 2014 December 1 (PID: GO-13743; PI: D.\ Thilker). 
To determine the precise position of the transient in the archival \textit{HST} frames, we registered our new WFC3/UVIS F555W observations of the active transient to the 2002 ACS/WFC F555W frames. Using centroid measurements of 22 stars in common between the two frames, we achieved an astrometric rms uncertainty of 0.03 WFC pixels ($0\farcs0015$). As shown in the rightmost column of Figure~\ref{fig:pre-imaging}, the transient is coincident with a red source detected in the 2002 ACS/WFC F814W frame but not in the F555W or F435W images. There is also an isolated source at the location in the WFC3/UVIS F218W and F225W images. We performed PSF-fitting photometry on the archival \textit{HST} images using DOLPHOT and report the results here on the Vega magnitude scale. The red source was detected by DOLPHOT as a ``good star'' at $\mathrm{F814W} = 23.29 \pm 0.02$ mag. For the UV source, we obtain $\mathrm{F218W} = 23.30\pm0.152$ and $\mathrm{F225W} = 23.348\pm0.059$\,mag. Given the presence of a red source, the UV measurements will be contaminated by the nonnegligible sensitivity of the UV filters beyond 0.4\,$\mu$m, referred to as ``red leak.''\footnote{The level of red leak affecting WFC3/UVIS filters is described in Section 6.5.2 of the WFC3 Instrument Handbook: \url{https://hst-docs.stsci.edu/wfc3ihb}} We describe our estimate of the level of this contamination and apply appropriate corrections in our analysis of the preexplosion SED below in Section~\ref{sec:UV_SED}. For the nondetections in F435W and F555W, we adopt $5\sigma$ limiting magnitudes based on the S/N reported by DOLPHOT of detected sources within 30 pixels ($0\farcs05$) of the transient location. This gives $\mathrm{F435W} > 27.5$\,mag and $\mathrm{F555W} > 27.5$\,mag.

We also searched for an IR counterpart to the precursor source in the available \textit{Spitzer} Super Mosaics in the IRAC [3.6], [4.5], [5.8], and [8.0] bands and the MIPS 24\,$\mu$m channel, each made from stacked images taken between 2003--2007. The [3.6] and [4.5] Super Mosaics were used as reference images for subtraction in our processing of the posteruption imaging as part of the SPIRITS program (see Section~\ref{sec:observations}). We identified a clear, point-like source in both of the [3.6] and [4.5] reference images, consistent with the location of the transient in our difference images (bottom center panel in Figure~\ref{fig:pre-imaging}). There is no preeruption counterpart detected at the location in the longer wavelength \textit{Spitzer} images. To obtain accurate photometry of the source and reliable upper limits we built source catalogs and performed PSF-fitting photometry for each of the IRAC Super Mosaic images using the DAOPHOT/ALLSTAR package \citep{stetson87}, where a model of the PSF is constructed using isolated stars in the image. Our PSF-fitting and photometry procedure, including corrections for the finite radius of the model PSF using the method of \citealp{khan17}, is described in detail in \citet{karambelkar19}. This gives $[3.6] = 18.76 \pm 0.09$ and $[4.5] = 18.33 \pm 0.07$\,mag (Vega) for the IR precursor source. For the longer wavelength IRAC images, we adopt $5\sigma$ limiting magnitudes based on the S/N of detected sources in our catalogs within $40\arcsec$ of the transient position and obtain $[5.8] > 17.8$ and $[8.0] > 16.8$\,mag. We then adopt a $5\sigma$ limiting magnitude of $11.2$\,mag based on the level of background variation near the transient position in the 24\,$\mu$m MIPS image.

\begin{figure}
\centering
\includegraphics[width=0.49\textwidth]{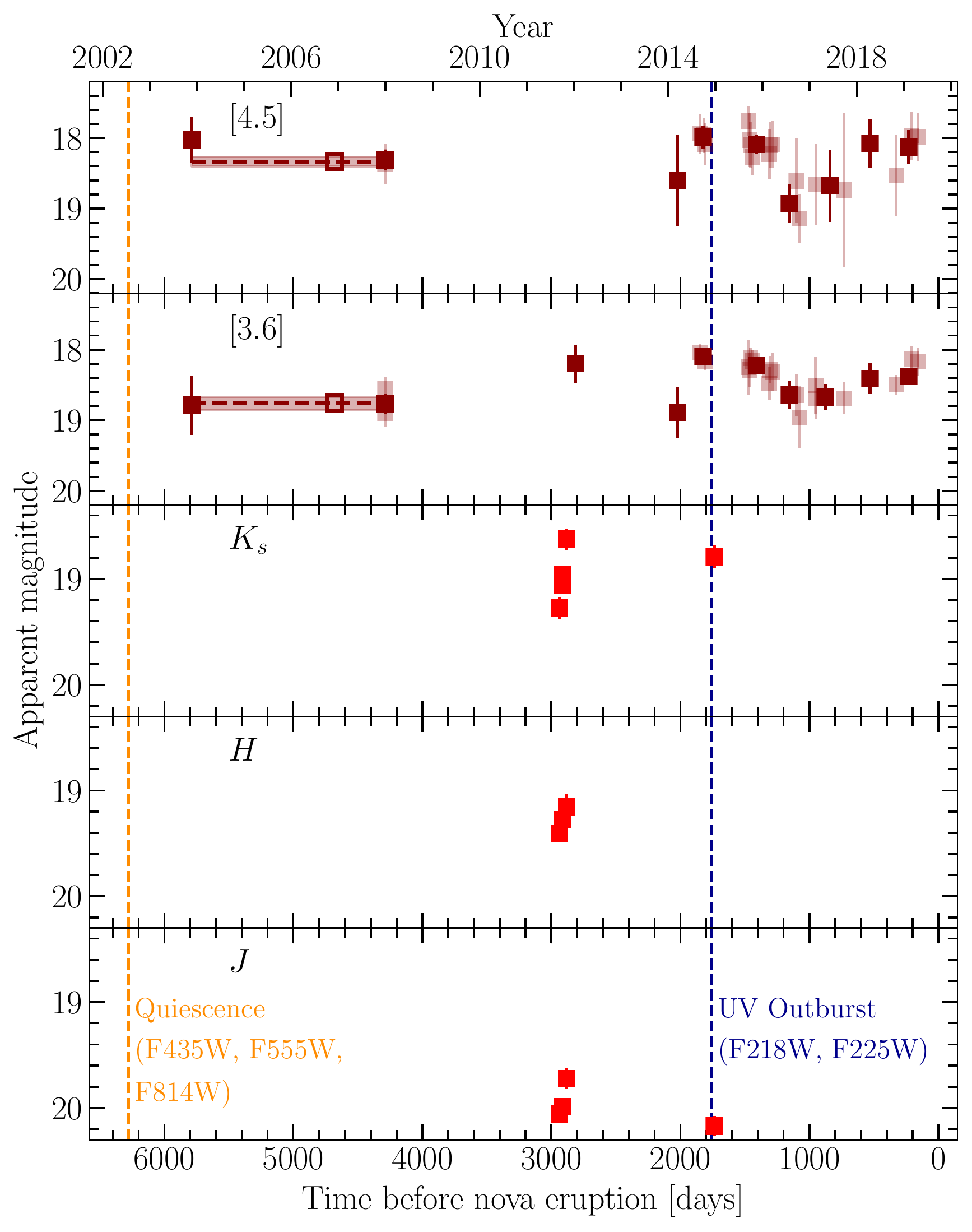}
\caption{\label{fig:pre_lcs}
Preexplosion light curves of the progenitor of AT\,2019qyl in the NIR $J$, $H$, and $Ks$ bands from Baade/FourStar and at [3.6] and [4.5] from \textit{Spitzer}/IRAC. Time on the $x$-axis is relative to the inferred eruption epoch of the nova at MJD 58752.12. As in Figure~\ref{fig:lcs}, all broadband photometry is corrected for Milky Way extinction. Our measurements of the IR counterpart source in the archival \textit{Spitzer}/IRAC [3.6] and [4.5] Super Mosaic images from PSF-fitting photometry are indicated by unfilled symbols, where the dashed horizontal lines and shaded bars indicate the range of observation dates included in the image stack and the uncertainty in the magnitude measurements, respectively. The individual epoch of NIR imaging at 55840.00 used in the construction of the precursor SED (see Figure~\ref{fig:pre_SED}) are indicated by black-outlined symbols. The times of the \textit{HST} ACS/WFC F453W, F555W, and F814W images (during quiescence) and the WFC3/UVIS F218W and F225W images (during outburst) are shown as the vertical dashed lines as labeled near the bottom of the figure. Colors used in this figure for different bands mirror those used in Figure~\ref{fig:pre_SED}.
}
\end{figure}

Finally, we examined preeruption $JHK_s$ images of NGC\,300 obtained with the FourStar IR camera \citep{persson13} on the Magellan Baade Telescope at LCO at five separate epochs between 2011 and 2014. The 2014 images were obtained as part of a concomitant monitoring program of SPIRITS galaxies. The IR precursor object is clearly detected in all three filters as a relatively isolated point source (see the top center panel of Figure~\ref{fig:pre-imaging}). We performed simple aperture photometry of the source with the aperture size set by the seeing in each image. The photometric zero points were calibrated using several isolated 2MASS stars in each image. Our preeruption, ground-based NIR photometry is provided in Table~\ref{table:pre-nearIR} and shown in Figure~\ref{fig:pre_lcs}.

\begin{deluxetable}{ccccc}
\tablecaption{Preeruption NIR Photometry from Baade/FourStar \label{table:pre-nearIR}}
\tablehead{\colhead{MJD} & \colhead{Phase\tablenotemark{a}} & \colhead{Band} & \colhead{App.\ Magnitude\tablenotemark{b}} \\ 
\colhead{} & \colhead{(days)} & \colhead{} & \colhead{(mag)} }
\startdata
55813.38 & -2938.74 & $H$              & $19.41$ $(0.08)$ \\
55814.30 & -2937.82 & $K_s$            & $19.3$ $(0.1)$   \\
55814.37 & -2937.75 & $J$              & $20.07$ $(0.09)$ \\
55839.39 & -2912.73 & $K_s$            & $19.06$ $(0.08)$ \\
55840.14 & -2911.98 & $K_s$            & $18.96$ $(0.08)$ \\
55840.20 & -2911.92 & $H$              & $19.28$ $(0.09)$ \\
55840.25 & -2911.87 & $J$              & $20.00$ $(0.06)$ \\
55869.21 & -2882.91 & $J$              & $19.7$ $(0.1)$   \\
55869.27 & -2882.85 & $H$              & $19.2$ $(0.1)$   \\
55869.31 & -2882.81 & $K_s$            & $18.6$ $(0.1)$   \\
57015.04 & -1737.08 & $K_s$            & $18.8$ $(0.1)$   \\
57015.06 & -1737.06 & $J$              & $20.2$ $(0.1)$   \\
\enddata
\tablenotetext{a}{Phase refers to time since $t_0$ on MJD 58752.12.}
\tablenotetext{b}{Vega magnitudes on the 2MASS system. 1$\sigma$ uncertainties are given in parentheses.}
\end{deluxetable}

\subsection{Preeruption Variability}\label{sec:pre-var}
We searched for historical variability of the preeruption source using the extensive archival coverage with \textit{Spitzer}/IRAC at [3.6] and [4.5] going back 16\,yr before the nova eruption and regular monitoring by SPIRITS since 2014. We obtained photometry on all difference images as described for the posteruption detections of the transients above in Section~\ref{sec:im_obs}. We then computed the Vega-system magnitudes of the source at each epoch by adding back in the reference flux measurements from our PSF-fitting photometry on the Super Mosaics. We further stacked individual measurements in bins of width $\Delta t = 300$~days to increase our S/N for detecting variability. Our preeruption \textit{Spitzer} photometry is provided in Table~\ref{table:pre-Spitzer} and shown in Figure~\ref{fig:pre_lcs} along with the multiepoch NIR photometry from Baade/FourStar.

\begin{deluxetable*}{cccccccc}
\tablecaption{Preeruption IR Photometry from \textit{Spitzer}/IRAC\tablenotemark{a}
\label{table:pre-Spitzer}}
\tablehead{\colhead{MJD} & \colhead{Phase\tablenotemark{b}} & 
\colhead{$dF_{\nu,[3.6]}$\tablenotemark{c}} & \colhead{$F_{\nu,[3.6]}$\tablenotemark{d}} & \colhead{$[3.6]$} &
\colhead{$dF_{\nu,[4.5]}$\tablenotemark{c}} & \colhead{$F_{\nu,[4.5]}$\tablenotemark{d}} & \colhead{$[4.5]$} \\ 
\colhead{} & \colhead{(days)} & \colhead{(mJy)} & \colhead{(mJy)} & \colhead{(mag)} &
\colhead{(mJy)} & \colhead{(mJy)} & \colhead{(mag)} }
\startdata
52964.06 & $-5788.06$ & $-0.2$ ($3.2$) & $ 8.6$ ($3.3$) & $ 18.8$ ($ 0.4$) & $ 2.2$ ($3.3$) & $10.6$ ($3.3$) & $ 18.0$ ($ 0.3$) \\
54462.88 & $-4289.24$ & $-1.1$ ($1.1$) & $ 7.7$ ($1.3$) & $ 18.9$ ($ 0.2$) & $-0.7$ ($1.9$) & $ 7.6$ ($2.0$) & $ 18.4$ ($ 0.3$) \\
54462.94 & $-4289.18$ & $ 1.8$ ($1.5$) & $10.6$ ($1.7$) & $ 18.6$ ($ 0.2$) & $-0.2$ ($0.9$) & $ 8.2$ ($1.0$) & $ 18.3$ ($ 0.1$) \\
55940.01 & $-2812.11$ & $ 5.9$ ($3.5$) & $14.7$ ($3.6$) & $ 18.2$ ($ 0.3$) & \nodata        & \nodata        & \nodata          \\
56729.51 & $-2022.61$ & $-1.0$ ($2.5$) & $ 7.8$ ($2.6$) & $ 18.9$ ($ 0.4$) & $-2.3$ ($3.8$) & $ 6.1$ ($3.8$) & $ 18.6$ ($ 0.6$) \\
56905.24 & $-1846.88$ & $ 8.3$ ($1.8$) & $17.1$ ($2.0$) & $ 18.0$ ($ 0.1$) & $ 3.2$ ($3.0$) & $11.6$ ($3.0$) & $ 17.9$ ($ 0.3$) \\
56933.42 & $-1818.71$ & $ 7.5$ ($1.2$) & $16.3$ ($1.4$) & $18.09$ ($0.09$) & $ 3.0$ ($2.6$) & $11.4$ ($2.6$) & $ 18.0$ ($ 0.2$) \\
56944.76 & $-1807.36$ & $ 6.3$ ($1.6$) & $15.1$ ($1.8$) & $ 18.2$ ($ 0.1$) & $ 1.6$ ($2.8$) & $10.0$ ($2.8$) & $ 18.1$ ($ 0.3$) \\
57279.43 & $-1472.69$ & $ 5.3$ ($5.0$) & $14.1$ ($5.0$) & $ 18.2$ ($ 0.4$) & $ 5.4$ ($2.6$) & $13.7$ ($2.6$) & $ 17.8$ ($ 0.2$) \\
57287.34 & $-1464.78$ & $ 4.8$ ($2.9$) & $13.6$ ($3.0$) & $ 18.3$ ($ 0.2$) & $ 2.2$ ($3.8$) & $10.6$ ($3.9$) & $ 18.0$ ($ 0.4$) \\
57294.71 & $-1457.41$ & $ 7.1$ ($1.8$) & $15.9$ ($2.0$) & $ 18.1$ ($ 0.1$) & $ 1.5$ ($3.0$) & $ 9.9$ ($3.1$) & $ 18.1$ ($ 0.3$) \\
57308.56 & $-1443.56$ & $ 6.5$ ($1.9$) & $15.3$ ($2.1$) & $ 18.2$ ($ 0.1$) & $ 0.0$ ($2.0$) & $ 8.4$ ($2.0$) & $ 18.3$ ($ 0.3$) \\
57440.10 & $-1312.02$ & $ 2.6$ ($2.4$) & $11.4$ ($2.5$) & $ 18.5$ ($ 0.2$) & $ 0.4$ ($2.8$) & $ 8.7$ ($2.9$) & $ 18.2$ ($ 0.3$) \\
57447.58 & $-1304.54$ & $ 4.7$ ($2.3$) & $13.5$ ($2.4$) & $ 18.3$ ($ 0.2$) & $ 1.6$ ($3.0$) & $ 9.9$ ($3.1$) & $ 18.1$ ($ 0.3$) \\
57466.94 & $-1285.18$ & $ 4.4$ ($3.2$) & $13.2$ ($3.3$) & $ 18.3$ ($ 0.3$) & $ 1.6$ ($3.0$) & $10.0$ ($3.1$) & $ 18.1$ ($ 0.3$) \\
57650.03 & $-1102.09$ & $ 1.0$ ($2.6$) & $ 9.8$ ($2.7$) & $ 18.6$ ($ 0.3$) & $-2.3$ ($3.5$) & $ 6.0$ ($3.5$) & $ 18.6$ ($ 0.6$) \\
57673.69 & $-1078.43$ & $-1.5$ ($2.9$) & $ 7.3$ ($3.0$) & $ 19.0$ ($ 0.4$) & $-4.8$ ($1.1$) & $ 3.5$ ($1.2$) & $ 19.1$ ($ 0.4$) \\
57800.65 & $ -951.47$ & $ 2.2$ ($4.0$) & $11.1$ ($4.1$) & $ 18.5$ ($ 0.4$) & \nodata        & \nodata        & \nodata          \\
57802.85 & $ -949.27$ & $ 0.6$ ($2.4$) & $ 9.4$ ($2.5$) & $ 18.7$ ($ 0.3$) & $-2.6$ ($3.1$) & $ 5.7$ ($3.2$) & $ 18.7$ ($ 0.6$) \\
58021.82 & $ -730.30$ & $ 0.6$ ($1.9$) & $ 9.4$ ($2.0$) & $ 18.7$ ($ 0.2$) & $-3.1$ ($5.7$) & $ 5.3$ ($5.7$) & $ 18.7$ ($ 1.1$) \\
58222.20 & $ -529.92$ & $ 3.3$ ($2.4$) & $12.1$ ($2.5$) & $ 18.4$ ($ 0.2$) & $ 1.7$ ($3.3$) & $10.1$ ($3.4$) & $ 18.1$ ($ 0.4$) \\
58425.21 & $ -326.91$ & $ 2.4$ ($1.3$) & $11.2$ ($1.5$) & $ 18.5$ ($ 0.1$) & $-1.9$ ($3.7$) & $ 6.5$ ($3.7$) & $ 18.5$ ($ 0.6$) \\
58545.83 & $ -206.29$ & $ 6.7$ ($2.6$) & $15.5$ ($2.7$) & $ 18.1$ ($ 0.2$) & $ 2.8$ ($3.5$) & $11.2$ ($3.6$) & $ 18.0$ ($ 0.3$) \\
58593.58 & $ -158.54$ & $ 6.3$ ($2.7$) & $15.1$ ($2.8$) & $ 18.2$ ($ 0.2$) & $ 2.7$ ($3.5$) & $11.0$ ($3.6$) & $ 18.0$ ($ 0.3$) \\
\enddata
\tablenotetext{a}{1$\sigma$ uncertainties given in parentheses.}
\tablenotetext{b}{Phase refers to time since $t_0$ on MJD 58752.12.}
\tablenotetext{c}{Measured flux in difference images.}
\tablenotetext{d}{Total flux, including that from PSF-fitting photometry on our reference images.}
\end{deluxetable*}

The [3.6] and [4.5] light curves appear to vary largely in sync, keeping a relatively constant color of $[3.6] - [4.5] \approx 0.2$\,mag. The full amplitude of variability observed in both bands is $\approx$0.8--0.9\,mag. Visually, we note a possible periodicity of around $\approx$1000--1300\,days based on the timing of the apparent light-curve peaks 2012 January ([3.6] only), 2014 September, and 2019 February. To test this, we fit the [3.6] and [4.5] light curves simultaneously using the \textsc{gatspy} \citep{vanderplas15a,vanderplas15b} implementation of the Lomb--Scargle method \citep{lomb76,scargle82}. We restricted our period search to sinusoidal signals given the sparse sampling of the light curves. The best-fitting period is $1820$\,days with a reduced $\chi^2$ value of $\chi^2/\nu = 0.7$. However, we note that this model is not a significant improvement over that of constant flux (no variability), which gives $\chi^2/\nu = 2.1$. Furthermore, the peak Lomb--Scargle power is only 0.66, equal to the power in a secondary period of $330$\,days. Thus, we are unable to make a robust determination of any periodicity with the available data.

Our temporal coverage in the NIR $JHK_s$ bands is more limited. During our 2011 coverage, we note a rise in all three bands over a period of 55\,days. The largest amplitude of variability is observed in the $K_s$ band to be 0.7\,mag; however, it is likely the $JHK_s$ amplitudes are larger than is represented in the available data. The timing of the 2014 UV imaging observations with \textit{HST} is also indicated in Figure~\ref{fig:pre_lcs}. Though we infer the precursor source must be in an outburst state based on the UV flux at this epoch (see Section~\ref{sec:UV_SED} below), we do not see a corresponding substantial increase in flux in the closest [3.6] and [4.5] images taken 48\,days earlier or the $JHK_s$ images taken 23\,days after. 

The observed IR variability may suggest semiregular pulsations of the AGB companion (Section~\ref{sec:IR_SED}). In this case, comparing the preeruption magnitudes of AT\,2019qyl to the [3.6] and [4.5] period-luminosity relations of \citet{whitelock17} and \cite{karambelkar19}, we would expect a period of $\approx$200--300\,days. This is somewhat shorter than the possible 330\,days period noted above. An alternative explanation may be orbital modulations owing to variable obscuration by an asymmetric or cone-shaped dust envelope. This has been seen in some symbiotic Miras (e.g., \citealp{whitelock87} and \citealp{munari90} for V407 Cyg), but may be unlikely for symbiotics with less-evolved donor stars, like AT2019qyl. Lastly, the observed brightening episodes may be associated with the long-wavelength tail emission from outbursts commonly observed in symbiotic systems, driven by a substantial increase in the luminosity of the hot component (see Section~\ref{sec:UV_SED}).

The location of AT\,2019qyl was also imaged in the optical at numerous epochs by ATLAS since October 2015, nearly four years before the nova eruption. We examined the available forced-photometry light curves for \replaced{ATLAS-c and ATLAS-o difference images}{difference images in the ATLAS-c and ATLAS-o observing bands}\footnote{ATLAS forced-photometry server: \url{https://fallingstar-data.com/forcedphot/}} \citep{tonry18,smith20} to constrain the optical variability of the progenitor during this time. In stacked measurements with 1 day binning, there are no prior detections of the source to typical 5$\sigma$ limiting magnitudes of $\gtrsim$19--20 mag in both bands. We computed Gaussian-weighted rolling sums of the S/N in the stacked data at the location of AT\,2019qyl and for eight nearby positions used as control light curves as described in \citet{rest21}. We find no evidence for significant prior optical variability at the location of AT\,2019qyl above the level of variation in the control light-curve sums. Unfortunately, the ATLAS light curves do not cover the time of the 2014 UV-bright outburst described below in Section~\ref{sec:UV_SED}.

\subsection{The Preeruption SED}\label{sec:pre-SED}
\begin{figure}
\centering
\includegraphics[width=0.49\textwidth]{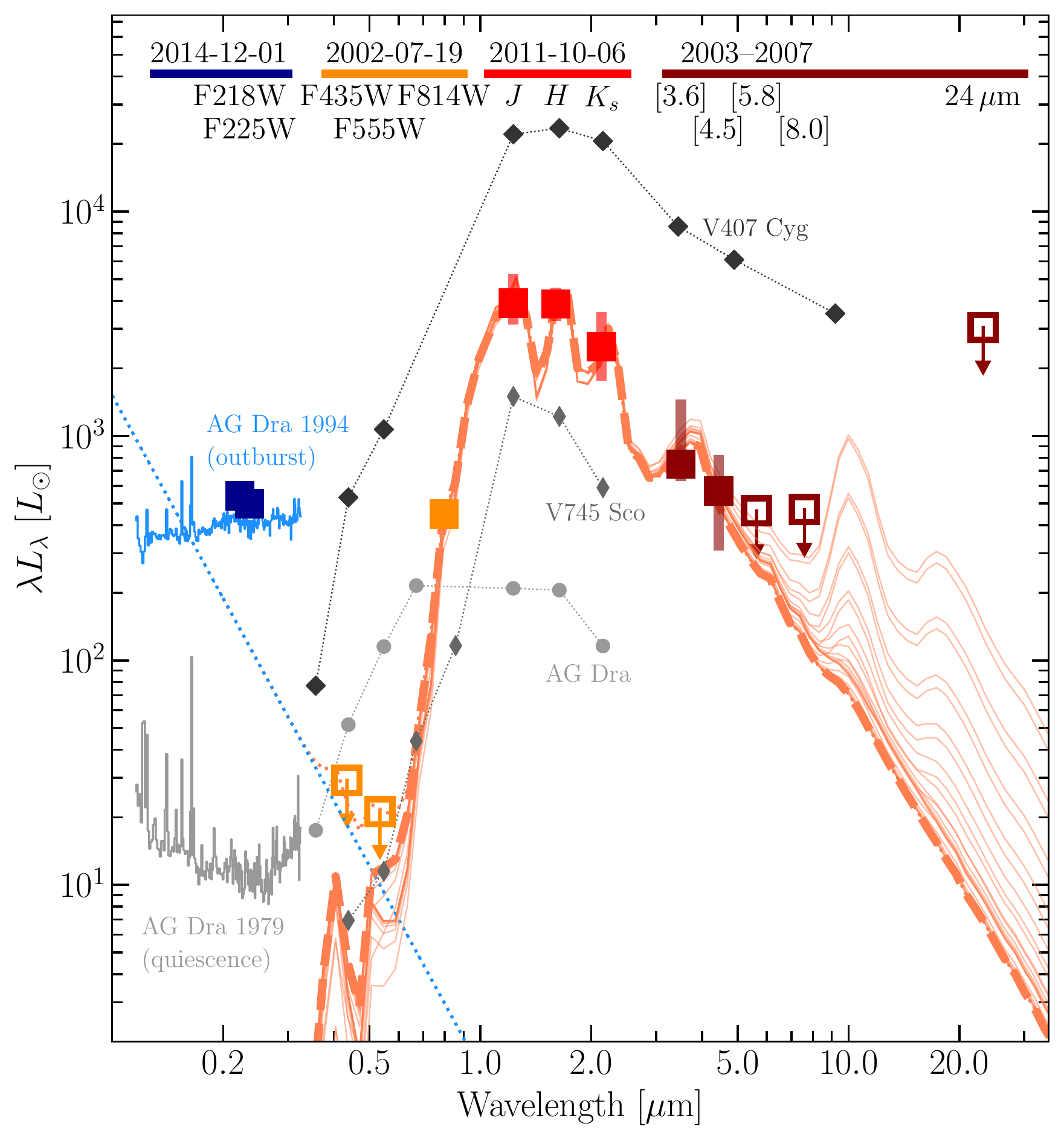}
\caption{\label{fig:pre_SED}
The multiepoch, preeruption SED of the AT\,2019qyl precursor from archival imaging is shown as the large squares. The time of observation for each point is indicated by different colors as labeled along the top of the figure. Error bars (not including the distance uncertainty) are smaller than the plotting symbols. $5\sigma$ upper limits from nondetections are indicated by unfilled squares with downward arrows. The level of variability observed in the preeruption light curves for a given band is indicated by the vertical bars (see Figure~\ref{fig:pre_lcs}). GRAMS O-rich AGB models that provide good fits to the optical and IR data (see main text) are shown as thin orange curves, and the best-fitting model for a star with $L_* = 3.7 \times 10^3$\,$L_{\odot}$ and $T_{\mathrm{eff}} = 2500$~K is indicated by the thick-lined, dashed curve. The maximum allowable flux for the Rayleigh-Jeans tail of a hot blackbody component (blue dotted curve) that is consistent with the visible \textit{HST} upper limits from 2002 (in sum with the best-fitting IR component model; orange dotted curve) is a factor of $\approx$4 below the observed UV fluxes from \textit{HST} in 2014, indicating a UV outburst at that epoch. For comparison, we show quiescent SEDs of the symbiotic binaries and novae V745~Sco (gray, thin diamonds; \citealp{schaefer10a,darnley12}) and V407~Cyg (dark gray, thick diamonds; \citealp{esipov88}), and that of the classical symbiotic star AG~Dra from photometry (light gray circles) along with IUE spectra from 1979 September 25 during quiescence (light gray) and 1994 July 28 during an outburst (blue; \citealp{skopal05}). For V407~Cyg we use an updated distance estimate of 3.6\,kpc based on the \citet{ita11} period-luminosity relation for HBB Miras with $K=3.3$\,mag from \citet{esipov88} and assuming $E(B-V) = 1$\,mag as in \citet{hachisu18}.
}
\end{figure}

Figure~\ref{fig:pre_SED} shows the multiepoch SED of the preeruption counterpart, constructed using the available archival imaging. We include UV, optical, and MIR measurements taken between 2002 and 2014 (\added{note that the measurements in different bands are not all contemporaneous;} see Section~\ref{sec:arc_im} for details). The photometric magnitudes were converted to band luminosities, $\lambda L_{\lambda}$, using the zero-point flux densities and effective wavelengths compiled by the Spanish Virtual Observatory (SVO) Filter Profile Service\footnote{Documentation for the SVO Filter Profile Service is available at \url{http://ivoa.net/documents/Notes/SVOFPSDAL/index.html} and \url{http://ivoa.net/documents/Notes/SVOFPSDAL/index.html}} for the appropriate filters \citep{svo1,svo2}. The most striking feature of the preeruption SED is the presence of distinct red and blue components, peaking in the IR and UV, respectively. The following sections provide an analysis of each individual component.

\subsubsection{The IR Component}\label{sec:IR_SED}
The IR SED component peaks in the $J$ band at a band luminosity of $\lambda L_{\lambda} = 3.9^{+1.1}_{-0.6} \times 10^3$\,$L_{\odot}$, where the uncertainty includes the observed variability. This suggests the presence of a cool giant, for which the broadband IR colors can be used to discriminate between AGB and RGB stars and provide a photometric classification \citep{whitelock87,blum06,cioni06,srinivasan09,whitelock09}. At $M_{K_s} = -7.3$\,mag the source is well above the tip of the RG branch and, following \citet{cioni06}, would be classified as an O-rich AGB star based on comparison to the 2MASS colors of AGB variables in the Large Magellanic Cloud (LMC). Based on the $J-[3.6]=1.2$ color, we can rule out a highly evolved ``superwind'' phase for an intermediate-mass AGB star ($J-[3.6]>3.1$), in which enhanced dust formation and dust-driven winds dramatically increase the mass-loss rate up to $10^{-4}~M_{\odot}$\,yr$^{-1}$ \citep{delfosse97,blum06,lagadec08,boyer15}. Thus, we infer that the IR component likely corresponds to a low-mass, O-rich AGB star in a pre-Mira phase of mass loss. The source has an intermediate luminosity between the symbiotic binary progenitors of the novae V745 Sco and V407 Cyg.

Based on the photometric classification, we fit the optical and IR SED of the precursor with the Grid of Red supergiant and Asymptotic Giant Branch ModelS (GRAMS; \citealp{sargent11,srinivasan11}) suite of radiative transfer models with O-rich (silicate) dust to estimate physical parameters of the star. This consists of a base grid of 1225 models of spherically symmetric shells of varying amounts of silicate dust \citep{ossenkopf92} around stars of constant mass-loss rates computed using the dust radiative code \texttt{2-Dust} \citep{ueta03}. The grid uses PHOENIX model photospheres \citep{kucinskas05,kucinskas06} for 1\,$M_{\odot}$ stars with effective temperatures, $T_{\mathrm{eff}}$, between 2100 and 4700\,K at fixed subsolar metallicity of $\log(Z/Z_{\odot}) = -0.5$ appropriate for the LMC and similarly for the environment of AT\,2019qyl inferred in Section~\ref{sec:host}. The surface gravity is fixed at $\log g = -0.5$, implying a stellar radius of $R_* \approx 295$\,$R_{\odot}$. The basic input parameters of each model are $T_{\mathrm{eff}}$, the inner radius of the dust shell, $R_{\mathrm{in}}$, and the optical depth of the dust shell at 10\,$\mu$m, $\tau_{10}$. The luminosity of the star, $L_{*}$, is computed by integrating the output spectrum over all wavelengths and can be scaled by a factor $s$ with the flux density of the output spectrum to match stars of varying brightness for a given $T_{\mathrm{eff}}$. For fixed $\log g$, the inferred stellar mass and radius then scale as $M_* \propto s$ and $R_* \propto s^{1/2}$. From $\tau_{10}$, the models also include the dust mass-loss rate, $\dot{M}_{\mathrm{dust}}$, assuming a dust wind speed of $v_w = 10$\,km\,s$^{-1}$.

In our fitting procedure, we include the F814W and $JHK_s$ SED points. For the IRAC points, we use the $[3.6] - [4.5]$ color as another constraint on the models owing to the large degree of variability observed in the [3.6] and [4.5] light curves at approximately constant color (see Section~\ref{sec:pre-var}). The base grid includes models with $R_{\mathrm{in}} = [3, 7, 11, 15] R_*$, but we found that the fits were largely insensitive to this parameter. Thus we restricted our fitting procedure to include only models with $R_{\mathrm{in}} = 11 R_*$ in order to reduce the number of free parameters. We compute the reduced-$\chi^2$ ($\chi^2/\nu$) statistic for each model and consider acceptable fits to be those within a factor of $e$ of the best-fitting model that minimizes $\chi^2/\nu$. We further rejected models that are inconsistent with any of the upper limits estimated from the F435W, F555W, [5.5], [8.0], and 24\,$\mu$m images. Acceptable model fits are shown along with the data in Figure~\ref{fig:pre_SED}. 

The best-fitting model has $T_{\mathrm{eff}} = 2500$\,K, $L_* =  3.7 \times 10^3$\,$L_{\odot}$, and $\tau_{10} = 10^{-4}$, and gives $\chi^2/\nu = 0.72$, indicating a very good fit to the data. From the range of acceptable models, these parameters are constrained to $2300 \leq T_{\mathrm{eff}}\ [\mathrm{K}] \leq 2500$, $3.7 \times 10^3 \leq L_*\ [L_{\odot}] \leq 4.3 \times 10^3$, and $10^{-4} \leq \tau_{10} \leq 0.21$, but we note that the allowed ranges include values near their respective minima in the grid for $T_{\mathrm{eff}}$ and $\tau_{10}$. Adopting the luminosity scaling relations for the fixed $\log g$ described above, the implied best-fitting values and allowed ranges of the stellar mass and radius are $M_* = 1.2$--$2.0~M_{\odot}$ and $R_* = 320$--$410~R_{\odot}$, with best-fitting values of $M_* = 1.2~M_{\odot}$ and $R_* = 320~R_{\odot}$, respectively. As is evident in Figure~\ref{fig:pre_SED}, we are limited in our ability to constrain the strength of the 10\,$\mu$m silicate feature owing to our lack of good photometric constraints in this wavelength region, permitting values of $\tau_{10}$ up to $0.21$. This corresponds to dust mass-loss rates between $\dot{M}_{\mathrm{dust}} = 2.9\times10^{-12}$ and $7.5\times10^{-9}\ M_{\odot}$\,yr$^{-1}$, and total wind mass-loss rates between $\dot{M_w} \approx 6 \times 10^{-10}$ and $2\times10^{-6}\ M_{\odot}$\,yr$^{-1}$ for a gas-to-dust ratio of 200 \citep{groenewegen09}.

We note that the assumption of spherical symmetry may be a poor approximation for the winds of symbiotic stars, where interactions with a massive WD can focus circumstellar material in the orbital plane of the binary (e.g., \citealp{booth16} for RS Oph). We discuss the possible implications of this for mass transfer in the progenitor of AT\,2019qyl further in Section~\ref{sec:bin_config}.

\subsubsection{The UV Component}\label{sec:UV_SED}
Given the deep limits from \textit{HST} at F435W and F555W, the bright UV counterpart detected in F218W and F225W in 2014 indicates a separate, hot component of the SED. These UV filters have nonnegligible out-of-band transmission in the red and IR, referred to as ``red leak.'' The IR source at the location will thus contaminate our UV flux measurements. We estimate the level of contamination assuming the best-fitting model spectrum for the IR source described above and compute the expected observed flux for this source through the WFC3/UVIS F218W and F225W filters using the STScI \textsc{pysynphot} package, which accounts for the full throughput calibration of \textit{HST} including every optical component. We estimate the IR source accounts for only 0.01\% and 0.02\% of the flux observed in F218W and F225W, respectively, and subtract these contributions from our estimate of the flux in each band. This contamination is thus very small and has a negligible effect on our analysis of the UV precursor source. 

At a UV band luminosity of $\lambda L_{\lambda} \approx 500$\,$L_{\odot}$, the 2014 UV source is too luminous to correspond to a quiescent phase spectrum consistent with the visible \textit{HST} F435W and F555W limits from 2002. We infer a limit on the UV luminosity of such a source by considering the maximum flux of the Rayleigh-Jeans tail of a hot blackbody ($f_\lambda \propto \lambda^{-4}$) that, when summed with our best-fitting spectrum for the IR source, is consistent with the 2002 \textit{HST} visible limits. As illustrated in Figure~\ref{fig:pre_SED}, the observed UV luminosity is a factor of $\approx$4 higher than this maximal estimate, implying that the precursor source must have been in an outburst state at the time of the 2014 UV observations. 

Classical symbiotic binaries are highly variable and are known to undergo frequent \added{accretion-powered outbursts}. Referred to as Z~And--type outbursts for the prototype system, these events typically correspond to an optical brightening by $\approx$2--3\,mag \added{and are distinct from the more luminous, thermonuclear-powered novae}. As shown in Figure~\ref{fig:pre_SED}, the observed 2014 UV luminosity for the AT\,2019qyl precursor shows good agreement with that of the classical symbiotic binary AG Dra observed during a 1994 outburst \citep{skopal05}, supporting this interpretation. As noted in Section~\ref{sec:pre-var}, the emission from the hot component during outburst can enhance the IR flux as well. Given the temporal coincidence between the 2014 UV observations and a peak in the MIR \textit{Spitzer} light curves (Figure~\ref{fig:pre_lcs}), it is plausible such outbursts may account for some portion of the observed IR variability. 


\section{Discussion}\label{sec:discussion}
In this study, we presented a detailed analysis of the early time evolution of a rapidly evolving nova and analyzed its symbiotic progenitor system. In this section, we first compare AT\,2019qyl with the sample of RS Oph--like Galactic RNe with RG companions and long orbital periods. We argue specifically that it is likely to be a short-recurrence-time system (Section~\ref{sec:RNe}). We then examine the properties of the outflow inferred from our unique data set for such an event, consisting of well-sampled early light curves and spectroscopy within the first day of the nova eruption. In particular, we discuss implications for possible outflow mechanisms and the importance of internal shocks as well as interactions between the ejecta and companion wind (Section~\ref{sec:shocks}). 

\subsection{AT\,2019qyl in the Context of RG RNe}\label{sec:RNe}
AT\,2019qyl is most similar to the subgroup of RNe with RG companions and orbital periods $\gtrsim$200\,days, including the famed RS Oph, as well as T CrB, V745 Sco, and V3890 Sgr \citep[e.g.,][ and references therein]{evans08,schaefer10a,darnley12}. \replaced{The 2010 nova outburst of V407 Cyg, which hosts an O-rich Mira companion, is also similar to this class in many respects, though it has not been confirmed to recur}{Other embedded RG novae, including the 2010 outburst of V407 Cyg, which hosts an O-rich Mira companion \citep[e.g.,][and references therein]{munari11,shore11,hinkle13}, and the peculiar 2015 outburst of V1535 Sco, which showed spectroscopic evidence of a high-velocity, bipolar outflow \citep{srivastava15,linford17}, bear similarities to this class, but are not yet known to recur}. Here we examine AT\,2019qyl in the context of this class in terms of the properties of the nova eruption itself and the binary configuration.

\subsubsection{Ejecta Velocity, Ejecta Mass, and WD Mass}
RG RNe are associated with very fast-evolving light curves, usually expected to correspond to the most massive WDs ($\gtrsim$1.2~$M_{\odot}$). Because of their larger surface gravities, they require low critical masses of accreted material to ignite the TNR, resulting in low ejecta masses of $\approx$10$^{-7}$--10$^{-6}~M_{\odot}$ and high velocities $\gtrsim$4000\,km\,s$^{-1}$ \citep[e.g.,][]{yaron05,hillman15,hillman16}. This appears to be borne out for RG RNe where ejecta and WD masses have been estimated in connection with very fast evolution timescales, including RS Oph and V745 Sco \citep{osborne11,page15}. AT\,2019qyl displays such high ejecta velocities (Section~\ref{sec:line_profiles}), and with $t_2(V) = 3.5$\,days and $t_3(V) = 10.3$\,days, it is very similar to those of RS~Oph, V745~Sco, and V3890~Sgr. In fact, AT\,2019qyl is one of only 17 known novae, including 15 Galactic systems along with the RN M31N~2008-12 and SMCN~2016-10a to have decline times $t_2 \lesssim 4$\,days \citep{hounsell10,page10,schaefer10a,munari11,orio15,darnley16,aydi18}. Utilizing the empirical relation presented by \citet{shara18}, we can estimate the mass of the ejected envelope as 

\begin{equation}
\log \left(\frac{M_{\mathrm{env}}}{M_{\odot}}\right)= 0.825 \log(t_2) - 6.18,
\end{equation}

\noindent from which we obtain $M_{\mathrm{env}} \approx 2\times10^{-6}~M_{\odot}$. As noted in Section~\ref{sec:lc_late}, the steep light-curve dropoff after $\approx$80\,days for AT\,2019qyl may be related to the end of steady nuclear burning on the WD surface, a proxy for $M_{\rm WD}$. Using Table~3 in \citet{hachisu06a}, we thereby estimate $M_{\rm WD}\approx1.2$--$1.3~M_{\odot}$. \citet{hachisu19} have found similarly massive WDs between $\approx$1.2 and 1.38~$M_{\odot}$ from optical light curves for RS Oph, T CrB, V745 Sco, and V407 Cyg.

\subsubsection{Companion Wind Interaction}
A fundamental difference between RG nova systems and CNe with main-sequence donors is that the explosion is embedded in the dense wind of the evolved companion, leading to distinct observational signatures. In particular, during the rise to maximum light for $t \lesssim 1$\,day, AT\,2019qyl shows narrow emission components of H and \ion{He}{1} ($\lesssim 300$~km\,s$^{-1}$; Figure~\ref{fig:line_profiles}) as well as numerous weaker narrow emission features, which likely arise from the slow-moving wind of the RG, flash ionized by the explosion on the WD. These features are common among RG novae observed early (e.g., \citealp{bode10} for RS Oph, \citealp{munari11} for V407 Cyg). As the nova evolves, shock interactions between the ejecta and the preexisting wind shape the spectra, producing broader emission components whose velocities decrease with time as the shock is decelerated by the dense wind, features seen clearly in the evolution of AT\,2019qyl (Section~\ref{sec:line_profiles}). 

\added{We can estimate the mass of swept-up CSM, $M_{\mathrm{CSM}}$, required to decelerate the shock by considering momentum conservation. We assume an ejecta mass of $2\times10^{-6}~M_{\odot}$ as estimated above and use the FWHM velocity of the H$\alpha$ intermediate-width emission component of 3500~km~s$^{-1}$ at $t=3.4$\,days and the FWHM of the Pa$\beta$ of line 350~km~s$^{-1}$ at $t=69.1$\,days as indicative of the shock velocity at those phases. We obtain an estimate of $M_{\mathrm{CSM}} \approx 1.8\times10^{-5}~M_{\odot}$. Assuming a power-law evolution of the shock velocity with time, also including the H$\alpha$ FWHM velocity of 1500~km~s$^{-1}$ at $t=7.9$\,days, we estimate a radius of the shock at $t=69.1$\,days of $4.2\times10^{14}$~cm. For a wind velocity of $v_w = 10$~km~s$^{-1}$, our estimate of $M_{\mathrm{CSM}}$ then corresponds to a mass-loss rate for the companion star of $\dot{M} \approx 1.3\times10^{-6}~M_{\odot}$~yr$^{-1}$. This is consistent with the upper end of the range mass-loss rates we estimate in Section~\ref{sec:IR_SED} based on the SED of the companion star. An obvious caveat here is that companion is not located at the center of the eruption, violating the assumption of spherical symmetry. The shock radius estimated here is seven times larger than the minimum orbital separation inferred for the binary in Section~\ref{sec:bin_config}. For larger binary separations, accounting for the effects of nonspherical CSM interaction could substantially alter these estimates.}

\added{Using the properties of the CSM inferred above, we can also estimate the expected CSM interaction luminosity. In the standard picture, the shock is slowed by the dense CSM resulting in the formation of a cold dense shell (CDS) behind the shock, where high densities and optical depths lead to the thermalization and rapid cooling of the shell via the radiation of continuum emission at optical wavelengths \citep[e.g.,][]{chugai94,chugai04,smith10a}. The maximum continuum luminosity from CSM interaction can then be estimated as 
\begin{equation}
L_{\mathrm{CSM}} = \frac{1}{2} \left(\frac{\dot{M}}{v_w}\right) v_{\mathrm{CDS}}^3, 
\end{equation}
\noindent where $v_{\mathrm{CDS}}$ is the velocity of the CDS. Taking the velocity of the broadened emission component observed at $t=3.4$~days for $v_{\mathrm{CDS}}$, we obtain
\begin{equation}
\begin{split}
    L_{\mathrm{CSM}} \approx &2.7 \times 10^{39} \left(\frac{\dot{M}}{10^{-6}~M_{\odot}~\mathrm{yr}^{-1}}\right) \left(\frac{v_w}{10~\mathrm{km~s}^{-1}}\right)^{-1} \\ 
    &\times \left(\frac{v_{\mathrm{CDS}}}{3500~\mathrm{km~s}^{-1}}\right)^{3}~\mathrm{erg}~\mathrm{s}^{-1}.
\end{split}
\end{equation}
}

\added{This value is comparable to the total luminosity of AT\,2019qyl inferred in Section~\ref{sec:SED} indicating that the interaction of the nova ejecta with the companion wind may provide a significant, or even dominant, portion of the UV-optical emission at early phases. We emphasize again that this is a maximal estimate---as some fraction of $L_{\mathrm{CSM}}$ may escape at other wavelengths, e.g. X-rays---for a highly idealized scenario. Especially at early phases, where the shock is still within the orbit of the binary, the effects of a nonspherical geometry will also be important. Finally, we expect that ongoing internal interactions between distinct, high-velocity ejecta components (see Section~\ref{sec:shocks}) could dampen the radiative output compared to interactions of the ejecta only with an external, slow-moving wind. 
}

These external interactions also generate nonthermal synchrotron as well as high-energy emission from hot plasma, observed in radio and X-rays for RS Oph for its 1985 and 2006 eruptions \citep[e.g.,][]{hjellming86,obrien06,sokoloski06}, and more recently for other embedded novae \citep[e.g.,][]{nelson12,delgado19,orio20}. V407 Cyg, in fact, was the first gamma-ray-detected nova \citep[e.g.,][]{abdo10}. The shocks were interpreted as arising from collisions between the nova ejecta and the dense Mira wind \citep{nelson12,martin13}, though \citet{martin18} argue that internal shocks could have given rise to the observed gamma-rays even without any contribution from ejecta--wind interactions (see Section~\ref{sec:shocks} below for a discussion of internal shocks for AT\,2019qyl). While secure radio or high-energy detections would be challenging to obtain at the distance of AT\,2019qyl in NGC\,300, the observed optical spectral signatures confirm the influence of the evolved, giant companion wind on the evolution of the eruption.

\subsubsection{The Binary Configuration}\label{sec:bin_config}
In Section~\ref{sec:IR_SED}, we estimated $R \approx 320~R_{\odot}$ and $M \approx 1.2~M_{\odot}$ for the AGB companion based on modeling of the IR component of the precursor source SED. For a binary with a similar mass WD (mass ratio $q \approx 1$), the ratio of the Roche radius, $R_{\mathrm{RL}}$ to the semimajor axis, $a$, is $0.38$ \citep{eggleton83}. Thus, we infer a lower limit of $a \gtrsim 840~R_{\odot}$ ($\gtrsim 4$~au) and an orbital period $P \gtrsim 1800$\,days ($\gtrsim 5$\, yr). This is several times longer than the orbital periods of the known Galactic RG RNe, which range between $\approx$200 and 500\,days. V407 Cyg, in contrast, is believed to have a very long orbital period of about 43~yr from observations of time-variable obscuration by a cone of dust \citep{munari90}.  

In addition to the orbital modulations, V407 Cyg is a hot-bottom-burning (HBB) Mira \citep{tatarnikova03} with a long period of 745 days and shows irregular optical brightenings ascribed to symbiotic outbursts like that seen for AT\,2019qyl in the UV in 2014 (Section~\ref{sec:UV_SED}). Interestingly, as noted in Section~\ref{sec:pre-var}, the IR precursor source of AT\,2019qyl shows a possible periodicity of $\approx$5\,yr---though we are unable to confirm this with currently available data---which is notably similar to that obtained by considering $R_{\mathrm{RL}}$ above. If the observed IR variability is taken to correspond to orbital modulations, we may thus expect the donor star to transfer mass onto the WD via Roche-lobe overflow. Alternatively, the IR variations could be attributed to semiregular pulsations of the low-mass AGB companion, allowing for a larger orbital separation and longer orbital period more similar to V407 Cyg. In this case, mass transfer onto the WD could still occur via a process called wind Roche-lobe overflow (WRLOF; \citealp{mohamed12}, and see \citealp{ilkiewicz19} for a direct application to V407 Cyg), or standard Bondi--Hoyle--Lyttleton (BHL) wind accretion \citep{hoyle39,bondi44}.

In the BHL scenario, assuming an orbital separation of 16~au similar to that inferred for V407 Cyg, we can estimate the wind-accretion rate of material onto the WD as it moves through the AGB wind following \citet{livio84}:

\begin{equation}\label{eq:Macc_wind}
\begin{split}
	\dot{M}_{\mathrm{acc}} = &1.8\times10^{-8} \left(\frac{M_{\mathrm{WD}}}{M_{\odot}} \right)^2 \left(\frac{v_{\mathrm{rel}}}{10~\mathrm{km~s}^{-1}} \right)^{-3} \\
	&\times \left(\frac{a}{10^{15}~\mathrm{cm}}\right)^{-2}
	\left(\frac{\dot{M}_w}{10^{-6}~M_{\odot}~\mathrm{yr}^{-1}}\right)^{2} \\
	&M_{\odot}~\mathrm{yr}^{-1},
\end{split}
\end{equation}

\noindent where $v_{\mathrm{rel}}$ is the relative velocity between the AGB wind and the WD given by $v_{\mathrm{rel}}^2 = v_w^2 + \left(\frac{2\pi a}{P}\right)^2$. Adopting $v_w = 10$\,km\,s$^{-1}$ and $\dot{M}_w \approx 10^{-6}~M_{\odot}$\,yr$^{-1}$ from our analysis of the IR source in Section~\ref{sec:IR_SED}, we obtain $\dot{M}_{\mathrm{acc}} \lesssim 1.3 \times 10^{-7}~M_{\odot}$\,yr$^{-1}$. For a $1.2~M_{\odot}$ WD, this would imply a recurrence time of $\sim 20$\,yr comparing to calculations by \citet{wolf13} and those presented in \citet{chomiuk20}. 

We find, however, that even at this wide separation, accretion onto the WD is likely to proceed via WRLOF. In this scenario, the AGB wind is concentrated into the orbital plane through interactions with the WD, leading to funneling of material and significantly enhanced accretion onto the WD. This process has been examined, for example, in both RS Oph \citep{booth16} and V407 Cyg \citep{ilkiewicz19}. Using 3D hydrodynamical simulations, \citet{mohamed12} found that WRLOF led to more than an order-of-magnitude increase in the accretion rate for a symbiotic Mira with a 20~au binary separation and an 0.6\,$M_{\odot}$ WD. For a massive WD like that in AT\,2019qyl, the interaction would be even stronger. Thus, the BHL accretion rate (and corresponding recurrence time) estimated above may only be a lower (upper) limit. We caution that there are substantial uncertainties in our knowledge of the relevant binary parameters for this system that could alter this estimate. Still, this suggests AT\,2019qyl is a strong candidate for a short-recurrence-time system similar to the Galactic RG RNe.



\subsection{Outflow Collisions and Internal Shocks in AT\,2019qyl}\label{sec:shocks}
In Section~\ref{sec:line_profiles}, we presented consistent evidence across several H and He lines for multiple, distinct velocity components in the nova outflow of AT\,2019qyl. We observe an early $\approx$2000~km\,s$^{-1}$ ``slow'' P Cygni absorption component at $t=0.17$--$0.27$\,days, which accelerates to $\approx$2500\,km\,s$^{-1}$ by the time of the optical peak at $0.97$\,days. Concurrently, we observe the appearance of a new ``fast'' absorption feature at $\approx$4000\,km\,s$^{-1}$, \added{which likely traces a separate component of the unshocked ejecta} and an emerging intermediate-width ($\approx$1300--2000~km\,s$^{-1}$) emission component. The broadened emission is indicative of shock interactions at this phase, likely between the early slow ejecta and the preexisting circumbinary material of the companion wind, but internal interactions between two ejecta components may also already contribute. At $t=3.37$\,days, the line profiles are in emission only and the intermediate-width component velocity has increased to $\approx$3500\,km\,s$^{-1}$, \replaced{suggesting continued and intensified shock interaction.}{indicative of the interaction of the fast ejecta with the earlier slow outflow and/or the ambient wind of the companion.} While we cannot completely disentangle the effects of internal ejection collisions and ejecta--wind interactions, the timing of the appearance of the broadened component is consistent with expectations for collisions within the observed multicomponent outflow. 

This is \replaced{similar to}{reminiscent of} the general picture of early evolution that has been recognized for several decades in CNe, but at notably higher velocities (see $100$--$1000$\,km\,s$^{-1}$ for the initial ``slow'' component; \citealp{mclaughlin42,gallagher78}). \citet{aydi20a} recently showed, with a sample of premaximum spectra of 12 novae, that this scenario may be universal---though their sample notably did not include any novae with decline times $t_2 < 6$\,days or embedded RG novae. 
In CNe, the superposition of the initial slow absorption component with the emerging broadened emission suggests that some portion of the ejecta must expand relatively freely, and consequently, that the outflows are aspherical. In one proposed scenario (discussed in more detail by, e.g., \citealp{aydi20a,chomiuk20}), the slow outflow arises when the accreted material is inflated by the underlying thermonuclear reactions and engulfs the binary in a common envelope. The outflow is then equatorially concentrated by the orbital motion of the binary. The fast outflow is established as a continuous, optically thick wind as in \citet{kato94}, driven by Fe opacity to the radiation of the ongoing burning on the WD surface \citep{friedjung66,bath76}.

For AT\,2019qyl, \added{the scenario is somewhat different. First, we do not see direct evidence for an aspherical outflow in that the slow absorption component does not persist at $t=3.37$\,days as a superimposed feature on the broadened emission. This may be attributable to the influence of the companion wind or point to an overall difference in the outflow geometry or ejection mechanisms.} Moreover, if we assume the ejecta velocities are comparable to their escape speeds ($v_{\mathrm{esc}} = [2GM/R_w]^{1/2}$), we can estimate the outflow launching radius, $R_w$, for a given component. For the $2000$\,km\,s$^{-1}$ ``slow'' component, we obtain $R_w \approx 8\times10^{9}$~cm ($0.1 R_{\odot}$, $8 R_{\mathrm{WD}}$). This is much smaller than the binary orbit estimated above, which excludes a common-envelope-like scenario in this case. An alternative scenario for the early mass ejection in AT\,2019qyl may be an impulsive ejection coincident with the TNR \citep{seaquist08,shore14,mason18}, or a short-duration wind powered by radioactive heating from $\beta$-decay \citep{starrfield08}. The opacity-driven, prolonged wind model of \citet{kato94} is likely still important given the sustained luminosity of novae for days after the TNR, as seen here for AT\,2019qyl. This may account for secondary phases of mass loss, i.e., the ``fast'' ejecta, and may also contribute to the observed acceleration of the slower ejecta (as in, e.g., \citealp{friedjung87}). 


\section{Summary}\label{sec:summary}
We have described the early discovery and prompt follow-up observations of AT~2019qyl, a very fast nova with an O-rich AGB donor in NGC~300. The DLT40 SN survey discovered AT~2019qyl within $\approx$1 day of eruption, quickly followed by multiband UV, optical, and IR imaging and a spectroscopic campaign that began just 2\,hr after discovery. These follow-up observations allowed us to constrain the eruption epoch and rise time in multiple bands, which occurred within one day, highlighting the need for a prompt, sustained response. Following the peaks, the optical light curves declined rapidly, fading by 2\,mag in $t_{2,V}=3.5$\,days in the $V$ band and placing AT\,2019qyl among the fastest known novae. The light curves decline smoothly until at least $t=71$\,days after which a steeper dropoff is observed, possibly corresponding to the end of nuclear burning on the WD surface. In analogy with similar systems, the rapid decline and timing of the dropoff point to a massive WD of $M_{\rm WD} \gtrsim 1.2~M_{\odot}$. The early evolution of the broadband SED is largely consistent with an expanding ``fireball'' that cools from $\approx$13000 to 8000\,K within the first day after the TNR. The source sustains a high, super-Eddington bolometric luminosity during this time of $1.5$--$2.3\times10^{39}$~erg~s$^{-1}$ ($8$--$13\times L_{\mathrm{Edd}}$). 

We obtained three high-quality optical spectra during the $\approx$1\,day rise to peak, which display a rich set of emission features consistent with a He/N spectral classification. In the earliest spectra, the strong H and He lines, in addition to numerous weaker emission features, show narrow profiles (limited by the resolution of our spectra of $\lesssim \mathrm{few} \times 100$~km\,s$^{-1}$), likely arising in the dense, preexisting wind of an RG companion star. The evolution of the P Cygni profiles of the strongest lines reveals multiple, distinct velocity components in the ejecta, including an early $2000$\,km\,s$^{-1}$ component that accelerates to $2500$\,km\,s$^{-1}$ concurrently with the appearance of a superimposed higher-velocity component at $4000$\,km\,s$^{-1}$ near the time of the optical peak around $t=1$\,day. By $t=3.37$\,days, the emission lines have broadened, indicative of ongoing shock interaction, the timing of which is consistent with internal collisions between the previously ejected outflows. The widths of the emission lines are then observed to decline, providing further evidence of the influence of the dense wind of an RG companion that acts to decelerate the ejecta. The early line evolution is remarkably similar to that long recognized in early optical spectra of CNe \citep{mclaughlin42,gallagher78}, and recently shown to be common, if not ubiquitous, in slower CN outbursts \citep{aydi20a}. In contrast with proposed scenarios for CNe, the high velocities in AT\,2019qyl require the outflows to originate close to the WD, possibly in a multiphase, opacity-driven wind, and preclude a common-envelope-like mass-loss mechanism. 

We also presented a detailed examination of the extensive archival data set available for this event, which reveals a red/IR- and UV-bright source at the nova location. Models of the preeruption SED suggest an O-rich AGB companion to the WD in a long-period binary ($\gtrsim$5\,yr). The IR source demonstrates significant variability over the course of 16~yr preceding the nova outburst, possibly consistent with pulsations of a semiregular AGB variable, orbital modulations, or symbiotic activity. The bright UV counterpart seen in 2014 \textit{HST} imaging also indicates a prior symbiotic-like outburst. Altogether, the properties of the outburst and progenitor system place AT\,2019qyl as a likely extragalactic analog to the RG RNe. Our early spectroscopic observations provide new, direct evidence of the importance of multicomponent outflows and internal collisions in contributing to the shock-powered emission observed in such systems. These results highlight the potential for high-cadence transient surveys with dedicated rapid-response capabilities to contribute to our understanding of some of the most common shock-powered transients in the universe. 

\acknowledgments
We thank the anonymous referee for their detailed and thoughtful comments, which helped us improve the clarity of this paper. We also thank E.\ Aydi for insightful discussions. We would like to thank Jorge Anais Vilchez, Abdo Campillay, Yilin Kong Riveros, and Natalie Ulloa for their help with Swope observations. We also thank the Magellan observers of Las Campanas Observatory and A.\ Monson for help with the Baade/FourStar data. 

Research by S.V.\ is supported by NSF grants AST–1813176 and AST-2008108. Support for {\it HST} program GO-15151 was provided by NASA through a grant from STScI. Research by D.J.S.\ is supported by NSF grants AST-1821967, AST-1821987, AST-1813708, AST-1813466, and AST-1908972, as well as by the Heising-Simons Foundation under grant \#2020-1864. H.E.B.\ and M.M.K.\ acknowledge support from Program number AR-15005, provided by NASA through grants from the Space Telescope Science Institute, which is operated by the Association of Universities for Research in Astronomy, Incorporated, under NASA contract NAS5-26555. P.A.W. and S.M. acknowledge funding from the South African National Research Foundation. The UCSC team is supported in part by NASA grant NNG17PX03C, NSF grant AST-1815935, the Gordon \& Betty Moore Foundation, the Heising-Simons Foundation, and by a fellowship from the David and Lucile Packard Foundation to R.J.F. R.D.G.\ was supported, in part, by the United States Air Force. D.A.C.\ acknowledges support from the National Science Foundation Graduate Research Fellowship under Grant DGE1339067. C.C.N.\ acknowledges support from the Ministry of Science and Technology (MoST) Taiwan under grant 104-2923-M-008-004-MY5. This work is part of the research programme VENI, with project number 016.192.277, which is (partly) financed by the Netherlands Organisation for Scientific Research (NWO). This project has received funding from the European Union's Horizon 2020 research and innovation programme under grant agreement No 730890. This material reflects only the authors' views and the Commission is not liable for any use that may be made of the information contained therein.

Based on observations obtained at the international Gemini Observatory (GS-2019B-Q-125), a program of NSF's NOIRLab, which is managed by the Association of Universities for Research in Astronomy (AURA) under a cooperative agreement with the National Science Foundation. on behalf of the Gemini Observatory partnership: the National Science Foundation (United States), National Research Council (Canada), Agencia Nacional de Investigaci\'{o}n y Desarrollo (Chile), Ministerio de Ciencia, Tecnolog\'{i}a e Innovaci\'{o}n (Argentina), Minist\'{e}rio da Ci\^{e}ncia, Tecnologia, Inova\c{c}\~{o}es e Comunica\c{c}\~{o}es (Brazil), and Korea Astronomy and Space Science Institute (Republic of Korea).  

Some of the observations reported in this paper were obtained with the Southern African Large Telescope (SALT). The SALT observations presented here were made through Rutgers University program 2019-1-MLT-004  (PI: Jha); this research at Rutgers is supported by NSF award AST-1615455.

This publication has made use of data collected at Lulin Observatory, partly supported by MoST grant 108-2112-M-008-001.

Some of the data presented herein were obtained at the W. M. Keck Observatory, which is operated as a scientific partnership among the California Institute of Technology, the University of California and the National Aeronautics and Space Administration. The Observatory was made possible by the generous financial support of the W. M. Keck Foundation.  The authors wish to recognize and acknowledge the very significant cultural role and reverence that the summit of Maunakea has always had within the indigenous Hawaiian community.  We are most fortunate to have the opportunity to conduct observations from this mountain.

This research is based on observations made with the NASA/ESA \textit{Hubble Space Telescope} obtained from the Space Telescope Science Institute, which is operated by the Association of Universities for Research in Astronomy, Inc., under NASA contract NAS 5–26555. These observations are associated with program(s) GO-15151, GO-9492, and GO-13743.

This work is based in part on archival data obtained with the {\it Spitzer Space Telescope}, which is operated by the Jet Propulsion Laboratory, California Institute of Technology, under a contract with NASA.

This research was made possible through the use of the AAVSO Photometric All-Sky Survey (APASS), funded by the Robert Martin Ayers Sciences Fund and NSF AST-1412587. This research has made use of the NASA/IPAC Extragalactic Database (NED), which is operated by the Jet Propulsion Laboratory, California Institute of Technology, under contract with the National Aeronautics and Space Administration. This research has made use of the SVO Filter Profile Service (\url{http://svo2.cab.inta-csic.es/theory/fps/}) supported from the Spanish MINECO through grant AYA2017-84089.

The national facility capability for SkyMapper has been funded through ARC LIEF grant LE130100104 from the Australian Research Council, awarded to the University of Sydney, the Australian National University, Swinburne University of Technology, the University of Queensland, the University of Western Australia, the University of Melbourne, Curtin University of Technology, Monash University and the Australian Astronomical Observatory. SkyMapper is owned and operated by The Australian National University's Research School of Astronomy and Astrophysics. The survey data were processed and provided by the SkyMapper Team at ANU. The SkyMapper node of the All-Sky Virtual Observatory (ASVO) is hosted at the National Computational Infrastructure (NCI). Development and support the SkyMapper node of the ASVO has been funded in part by Astronomy Australia Limited (AAL) and the Australian Government through the Commonwealth's Education Investment Fund (EIF) and National Collaborative Research Infrastructure Strategy (NCRIS), particularly the National eResearch Collaboration Tools and Resources (NeCTAR) and the Australian National Data Service Projects (ANDS).

\facility{CTIO:PROMPT, LCOGT (Sinistro), Swope (CCD),  LO:1m, Keck:II (NIRES), Gemini:South (GMOS), SALT (RSS), FTN (FLOYDS), NOT (ALFOSC), HST (WFC3), Spitzer (IRAC), Swift (UVOT), Magellan:Baade (FourStar)}

\software{AstroDrizzle \citep[][\url{http://drizzlepac.stsci.edu}]{hack12}, DOLPHOT \citep{dolphin00,dolphin16}, IRAF \citep{tody86,tody93}, Gemini IRAF package (\url{http://www.gemini.edu/sciops/data-and-results/processing-software}), PyRAF (\url{http://www.stsci.edu/institute/software_hardware/pyraf}), Spextool \citep{cushing04,vacca03}, PypeIt \citep[][\url{https://pypeit.readthedocs.io/en/latest/}]{prochaska20,pypeit_zenodo}, BANZAI \citep[][\url{https://github.com/LCOGT/banzai}]{mccully18}, lcogtsnpipe \citep[][\url{https://github.com/LCOGT/lcogtsnpipe}]{valenti16}, photpipe \citep[][]{rest05}, HEAsoft (\url{https://heasarc.gsfc.nasa.gov/docs/software/heasoft/}), pysynphot (\url{https://pysynphot.readthedocs.io/en/latest/}), BBFit (\url{https://github.com/nblago/utils/blob/master/src/model/BBFit.py}), emcee \citep[][\url{https://emcee.readthedocs.io/en/stable/}]{foreman-mackey13}
}


\bibliography{jencson}{}
\bibliographystyle{aasjournal}



\end{document}